\documentclass[aps,prb,twocolumn,superscriptaddress]{revtex4}

\usepackage{makeidx}
\usepackage{amsmath}
\usepackage{amsthm}
\usepackage{amssymb}
\usepackage{amsxtra}
\usepackage{amscd}
\usepackage[mathscr]{eucal}
\usepackage{boxedminipage}
\usepackage{bbm}
\usepackage{epsfig}
\usepackage{epic,eepic}
\usepackage{latexsym}
\usepackage{graphics}

\newcommand{\bra}[1]{\ensuremath{\langle#1|}}
\newcommand{\ket}[1]{\ensuremath{|#1\rangle}}
\newcommand{\be}{\begin{equation}}
\newcommand{\ee}{\end{equation}}

\newcommand{\olcite}[1]{[\onlinecite{#1}]}

\newcommand{\Eq}[1]{Eq.~(\ref{#1})}
\newcommand{\eq}[1]{Eq.~(\ref{#1})}

\newcommand{\Fig}[1]{Fig.~\ref{#1}}

\newcommand{\abs}[1]{\ensuremath{\left|#1\right|}}

\newcommand{\aver}[1]{\ensuremath{\big<#1 \big>}}

\newcommand{\pdag}{{\phantom{\dagger}}}

\renewcommand{\Im}{\ensuremath{\mathrm{Im}\,}}

\newcommand{\ini}{\ensuremath{\ket{\uparrow \dots \uparrow \downarrow
      \dots \downarrow}}}
\newcommand{\ketini}{\ensuremath{\ket{\mathrm{ini}}}}
\newcommand{\init}{\ensuremath{\mathrm{ini}}}
\begin{document}

\title{Real-time dynamics in spin-$1/2$ chains 
with adaptive time-dependent DMRG}
\author{Dominique Gobert}
\affiliation{Institute for Theoretical Physics C, RWTH Aachen, D-52056 Aachen, Germany}
\affiliation{Physics Department and CeNS, LMU M\"unchen, Theresienstr.\ 37, D-80333 M\"unchen, Germany}
\author{Corinna Kollath}
\affiliation{Institute for Theoretical Physics C, RWTH Aachen, D-52056 Aachen, Germany}
\affiliation{Physics Department and CeNS, LMU M\"unchen, Theresienstr.\ 37, D-80333 M\"unchen, Germany}
\author{Ulrich Schollw\"{o}ck}
\affiliation{Institute for Theoretical Physics C, RWTH Aachen, D-52056 Aachen, Germany}
\author{Gunter Sch\"{u}tz}
\affiliation{Institut f\"{u}r Festk\"{o}rperforschung, Forschungszentrum
J\"{u}lich, D-52425 J\"{u}lich, Germany}
\date{\today}

\begin{abstract}
We investigate the influence of different interaction strengths and
dimerizations on the magnetization transport in antiferromagnetic 
spin $1/2$ XXZ-chains.
We focus on the real-time evolution of the inhomogeneous 
initial state $\ini$ in
using the
adaptive time-dependent density-matrix renormalization group 
(adaptive t-DMRG). 
Time-scales accessible to us are of the order of 100 units of time
measured in $\hbar/J$ for almost negligible error in the
observables. We find ballistic magnetization transport for small
$S^zS^z$-interaction and arbitrary dimerization, but almost no transport for
stronger $S^zS^z$-interaction, with a sharp crossover at $J^z=1$. 
% At $J^z=1$ results indicate superdiffusive transport. 
Additionally, we perform a detailed analysis of the 
error made by the adaptive time-dependent
DMRG using the fact that the evolution in the $XX$-model is known exactly. We
find that the
error at small times is dominated by the error made by the Trotter
decomposition 
whereas for longer times the DMRG truncation error becomes the most important,
with a very sharp crossover at some ``runaway'' time. Overall, errors
are extremely small before the ``runaway'' time. 
\end{abstract}


\maketitle

\section{\label{time_intro} Introduction}

The transport properties of spin chains have attracted much attention recently,
not only due to the possible applications to information storage,
spintronics, and quantum information processing, but also because
they allow to study general aspects of nonequilibrium dynamics in a
comparably simple system.
Nonequilibrium phenomena are a vast and despite all progress 
still poorly understood 
field of statistical physics. It is therefore
useful to have a simple model at hand that allows to study general
questions rather explicitly. 
In order to study nonequilibrium phenomena, a real-time description
is particularly intuitive and useful. In this paper, we study
the time-evolution of a spin-$1/2$ chain by solving the full many-body
Schr\"odinger equation.

Recently, new developments in the area of non-equilibrium
physics were stimulated by the experimental progress in the field of
ultracold atoms. These systems have the advantage that their
parameters can be tuned in time with high accuracy and on very short
time-scales. 
It was proposed that spin-$1/2$ chains can be realized in these
systems as well 
\cite{KuklovSvistunov2003, DuanLukin2003, AltmanLukin2003, MandelBloch2003}, 
namely as a mixture of atoms of two species, say $A$ and $B$.
If these atoms are studied in an optical lattice 
with an average filling of one atom
per site and with a very strong repulsive interaction between the
atoms, such that multiple occupancy is suppressed, the system can be
mapped onto an effective spin-$1/2$ model. 
In this effective model the state with atom
$A$ occupying a given lattice site corresponds to, say,
$\uparrow$, and likewise $B$ to $\downarrow$.

In this paper, we study the time evolution of an initial state 
$\ini$ (or $\ket{A \cdots AB \cdots B}$),
i.e.\ with all spins on the left half pointing up along the z axis,
and all spins on the right half pointing down, under the effect of
a nearest-neighbour spin interaction (see Eq. \ref{Hamilton}). 
This system can also be interpreted as an oversimplified picture for spin
transport between two coupled reservoirs of completely polarized spins of
opposite direction in the two reservoirs. 
We are mainly interested in the following questions: 
Does the state evolve into a simple long time limit?
If so, how is this limit reached? 
On what properties does the long time behaviour depend?

Analytical results for this problem are essentially restricted to the
$XX$-chain which is amenable to an exact solution
\cite{AntalSchutz99}.
Here, a scaling relation for the long-time limit was found.
However, it is presently not known whether this relation 
is general, or whether
it relies on special properties of the XX model.
If a long-time limit exists for other models as well, 
the question arises which of its characteristics are universal,
and which depend on certain system properties.

Directly solving the time-dependent Schr\"odinger 
equation for interacting many-body
systems is highly nontrivial. 
A recently developed numerical method, the adaptive time-dependent DMRG
\cite{Vidal04, DaleyVidal04,WhiteFeiguin04} (adaptive t-DMRG), 
enables us to perform this task.
The two main conditions for this method to be applicable, namely that
the system must be one-dimensional and have nearest-neighbour
interactions only, are met for the present model. 
Efforts to generalize the DMRG method to time-depending problems
relaxing these constraints are under way \cite{WhiteFeiguin04b}.

As so far no detailed error analysis of this new method has been
performed, an important aspect of the present work is that
besides their own physical interest, spin-$1/2$ chains provide
an excellent benchmark for the adaptive time-dependent DMRG,
because of the nontrivial exact solution 
for the XX model, against which the method can be compared.
This allows us to analyze the accuracy of the adaptive time-dependent DMRG very
explicitly, namely to address the questions what kinds of errors can
occur in principle, which ones of these dominate in practice, and how
they can be minimized. We find that the time-scales accessible to us
are about $100 \hbar/J$ with an neglegible error in the observables at
very moderate numerical cost.

The outline of our paper is as follows: In section \ref{time_model} we
introduce the model and its characteristics. In section
\ref{time_method} we summarize the method and in section
\ref{time_error} a detailed error analysis is performed. These two
sections may be skipped by readers mainly interested in the physics
and not in the details of the method. In section \ref{time_scaling}
we present our results for the long time limit of the time
evolution for different interaction and dimerization strength.

\section{\label{time_model} Model and initial state}
In this paper we analyze the dynamics of the 
inhomogeneous initial state $\ket{\init}=\ini$ 
on the one-dimensional spin-$1/2$ chains with
interactions given by the Heisenberg model
\be
\label{Hamilton}
H = \sum_n J_n (S^x_n S^x_{n+1} + S^y_n S^y_{n+1} + J_z S^z_n S^z_{n+1})
\equiv \sum_n h_n.
\ee
Here, $\vec{S}_n$ is the spin operator on site $n$, and $J_n$, $J_z$
are interaction constants.
We consider dimerized models where
$J^z=const$ and $J_n = ( 1 + (-1)^n \delta)$, $\delta$ being the dimerization
coefficient. 
For $\delta > 0$, 
the ``strong bond'' with $J_n = 1+\delta$ is chosen to be at
the center, where the spin flip of the initial state is located.

We have chosen our energy unit such that $J_n=1$ for the homogeneous case 
$\delta = 0$. We also set $\hbar = 1$, defining time to be 1/energy with
the energy unit chosen as just mentioned. 

\begin{figure}
\epsfig{file=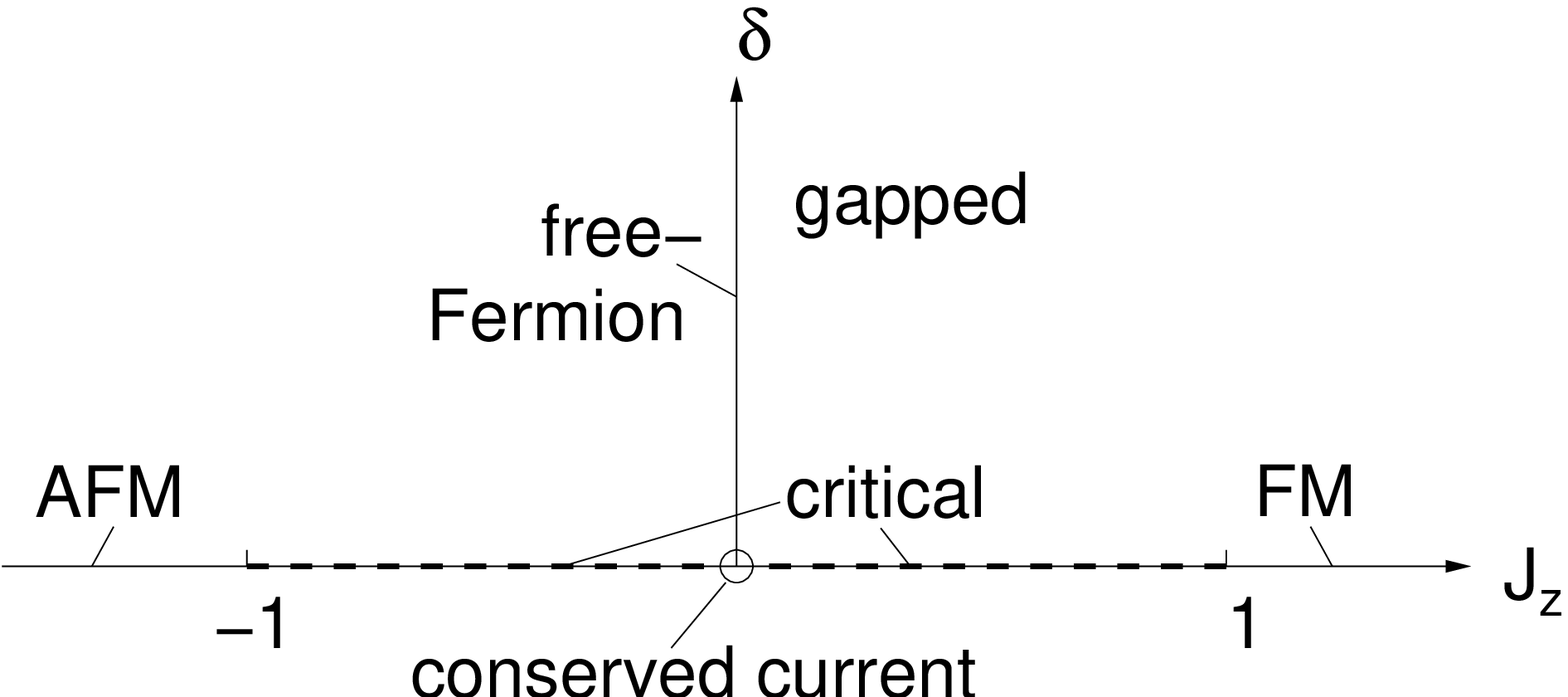, width=.9\linewidth}
 \caption[Quantum phase diagram of the Heisenberg model.]{ \label{phasediagram}
Quantum phase diagram of the Heisenberg model, \eq{Hamilton}. 
See \olcite{MikeskaKolezhuk04, Schuetz94} for details.
}
\end{figure}
The quantum phase diagram of this model at zero temperature is well
known (see \cite{MikeskaKolezhuk04,Schuetz94})
and sketched in \Fig{phasediagram}. 
For the homogeneous case, $\delta=0$, the ground state has ferromagnetic (FM) /
antiferromagnetic (AFM) order with a gap in the excitation 
spectrum for $J_z<-1$ and $J_z>1$,
respectively. The gap closes if $|J_z|$
approaches $1$ from above, and the model becomes critical for $-1<J_z<1$,
i.e.\ gapless in the thermodynamic limit, with correlation functions
showing a power-law decay. 
The model at the point  $J_z = \delta = 0$ is known as the
XX model. It has the special property that the spin-current operator
$J =\sum_n j_n$ is conserved, i.e. $[J, H]=0$. Here $j_n= J_n \Im
(S_n^+ S_{n+1}^-)$ is the current operator 
on the bond between site $n$ and $n+1$.
For finite dimerization, $\delta \neq 0$, the spectrum is again gapped
for all values of $J_z$.

Often it is useful to map the Heisenberg model
onto a model of spinless fermions:
%''and references therein'' -- see .bib file
\begin{eqnarray}
H&= &\sum_n J_n \left[ \frac{1}{2} ( c_n^\dagger c_{n+1}^\pdag +
  c_{n+1}^\dagger c_n^\pdag)\right. \nonumber \\
&&\left. + J_z
(c^\dagger_n c_n^\pdag- \frac{1}{2}) (c^\dagger_{n+1} c^\pdag_{n+1}- \frac{1}{2})\right]. 
\end{eqnarray}
In this picture, the first two terms in \eq{Hamilton} describe
nearest-neighbour hopping, whereas the third term (the one proportional
to $J_z$) describes a
density-density interaction between nearest neighbours. In particular,
the case $J_z=0$ describes free fermions on a lattice, and can 
be solved exactly \cite{LiebMattis1961}.

The time-evolution
under the influence of a time-independent 
Hamiltonian $H$ as in \Eq{Hamilton} is given by:
\be
\label{time_evo}
\ket{\psi(t)} = U(t) \ketini \;\;\;\; \mathrm{with} \; 
U(t) = \exp (- i  H t).
\ee

In most of the phases shown in \Fig{phasediagram}, the state
 $\ketini = \ini$ contains many high-energy excitations and is
 thus far from equilibrium.
In the following, we briefly discuss these phases separately.
% The nontrivial dynamics in the system stems from the fact that
% $\ketini$ is \emph{not} an eigenstate in most of the phases. 
\\ -- Deep in the ferromagnetic phase, $J_z < -1$,
 $\ketini$ corresponds to a state with one domain wall
 between the two degenerate ground states. 
 For  $J_z \to -\infty$ it
 is identical to the ground state (with boundary conditions given by $\ket{\uparrow}$ and
 $\ket{\downarrow}$ and $S_z^{\textrm{tot}}=0$), and therefore stationary.
 For finite $J_z$, it is no longer identical to the
 ground state, but still close to it \cite{Mikeska91}.
%\new{look up energy spectrum!!!}   
% with no energy  dissipation. 
\\ -- In the antiferromagnetic phase, $J_z > 1$, the state $\ketini$ is highly
 excited. One could view it as a state with almost the maximum number
 of domain walls of staggered magnetization. 

In this context, it is interesting to note that the sign of $J_z$ does
not matter for the time evolution of physical quantities, as long as the
initial state is described by a purely real wave function (which is
the case for our choice of $\ketini$), 
since the sign change in $J_z$ can be compensated by a gauge
transformation that inverts the sign of the hopping terms $S^x S^x$,
$S^yS^y$ in \Eq{Hamilton}, plus a complex conjugation of \Eq{time_evo}.
In particular, the time-evolution of the low-energy one domain-wall state in
 the FM is the same as the evolution of the high-energy many
 domain-walls state in the AFM. 
We therefore restrict ourselves to the case $J_z>0$.
%, as the results allow the direct analysis of the $J_z<0$ case.
\\ -- In the
 critical phase $\delta=0$ and $|J_z|<1$,
the ground state is a state with power-law correlations in the $xy$-plane. 
Here, the state $\ketini$ is not close to any particular
 eigenstate of the system, but 
contains many excited states throughout the energy spectrum, depending
 on the value of $J_z$:
The energy expectation value of $\ketini$ is low as 
$J^z\rightarrow -1$ and high as $J^z\rightarrow 1$. 

The time evolution
delocalizes the domain wall over the entire chain. For $J_z=0$, the      
time-evolution of the system can be solved exactly. For example, the
magnetization profile for the initial state $\ketini$ reads 
 \cite{AntalSchutz99}: 
\be
\label{Sz_nt}
S_z(n,t) = \bra{\psi(t)} S_n^z \ket{\psi(t)}
 =  -1/2\sum_{j=1-n}^{n-1}J_j^2(t),
\ee
where $J_j$ is the Bessel function of the first kind. 
$n=\ldots,-3,-2,-1,0,1,2,3,\ldots$ labels chain sites with the
convention that the first site in the right half of the chain has label 
$n=1$.
\\ -- In the dimerized phase, $\delta \neq 0$, the mentioned characteristics
 remain unchanged.
However, here the delocalization becomes confined to pairs of
 neighbouring sites in the limit $\delta \to 1$. 

We finally note that the total
energy and magnetization of the system are conserved at all times,
such that even for long times the state cannot relax to the ground
state.

\section{\label{time_method} Outline of the adaptive time-dependent 
DMRG for spin chains}

In order to determine the time-evolution of \eq{time_evo}, we use the
adaptive t-DMRG method \cite{DaleyVidal04,WhiteFeiguin04},
which has been introduced as an extension of 
standard DMRG using the TEBD algorithm of Vidal \cite{Vidal04}. 
It allows to evaluate the time-evolution for 
one-dimensional quantum chains with nearest-neighbour (possibly time-dependent)
interactions. In this paper, we consider the
case of a time-independent Hamiltonian where the dynamics is
introduced by a nonequilibrium initial state at $t=0$. To set the stage
for the error analysis, we briefly review adaptive t-DMRG, assuming the reader 
to be familiar with standard static zero temperature DMRG (see, e.g.\ 
\olcite{WhiteNoack99,Schollwoeck04}).

\begin{figure}
\epsfig{file=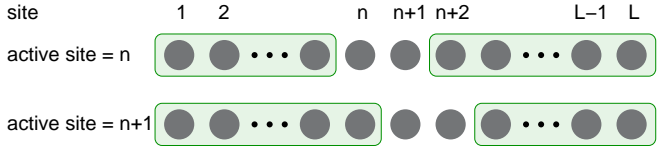, width=1\linewidth}
 \caption{ \label{DMRG_basis}
Illustration of the DMRG bases with active site $n$ and $n+1$, respectively.
}
\end{figure}

In the standard finite-system DMRG algorithm, a quantum-mechanical state
on a one-dimensional chain with $L$ sites is represented in a
particular tensor product basis, namely as
\be
\label{time_DMRG_state}
\ket{\psi} = \!\!\!
\sum_{\alpha \sigma \tau \beta}  \!\!
\psi_{\alpha \sigma \tau \beta} 
\ket{\alpha}_{1..n-1} \ket{\sigma}_n \ket{\tau}_{n+1} \ket{\beta}_{n+2..L}
\ee
as illustrated in the upper part of \Fig{DMRG_basis}.
Here, $\ket{\sigma}_n$, $\ket{\tau}_{n+1}$ are
complete bases on sites $n$, $n+1$; $\ket{\alpha}_{1..n-1}$
and  $\ket{\beta}_{n+2..L}$ are states on the subchains with sites $1,...,n-1$
and $n+2,...,L$, respectively. The states $\ket{\alpha}_{1..n-1}$ and
$\ket{\beta}_{n+2..L}$ form truncated bases, i.e. they do not span
the full Hilbert space on their respective subchains, but only a
subspace of dimension $m$, chosen to allow an optimal 
approximation of the true physical state. 
% under this constraint.
In the representation of \eq{time_DMRG_state}, we call site $n$ the
``active site''. The algorithm now consists of moving (``sweeping'') 
the position of the active site several times from the left to the right 
end of the chain and back, and constructing optimized truncated bases for
the subchains.

A DMRG step during such a sweep, say, to the right now consists of a
basis transformation from the old (truncated) basis 
$\ket{\alpha}_{1..n-1} \ket{\sigma}_n \ket{\tau}_{n+1} \ket{\beta}_{n+2..L}$ with
active site $n$ to a new one
$\ket{\alpha'}_{1..n} \ket{\sigma'}_{n+1} \ket{\tau'}_{n+2} \ket{\beta'}_{n+3..L}$
with active site $n+1$ as shown in \Fig{DMRG_basis}.
The states $\ket{\alpha'}_{1..n}$ representing the sites $1,...,n$ are
linear combinations of the old basis vectors $\ket{\alpha}_{1..n-1}\ket{\sigma}_n$.
Not all linear combinations are kept because of the DMRG truncation
that limits the number of states $\ket{\alpha'}_{1..n}$ to $m$ states.
For this reason, the state $\ket{\psi}$ can in general be represented
in the new basis only up to some truncation error.
The DMRG truncation algorithm (described in
\olcite{WhiteNoack99,Schollwoeck04}) provides a
unique optimal choice for the states $\ket{\alpha'}$ that minimizes this error
(which is then typically as low  as $10^{-10}$ or so)
and thus allows for the optimal representation of particular ``target'' states.
The basis vectors $\ket{\beta'}_{n+3..L}$ are taken from  stored
values from the
previous sweep to the left.
-- A sweep to the left (i.e. from active site $n$ to $n-1$)
works in the same way, with the role of 
$\ket{\alpha'}$ and  $\ket{\beta'}$ interchanged.

In standard DMRG, a mere transformation of the state
$\ket{\psi}$ from one basis to the other -- known as White's state
prediction \cite{White1996}-- is possible and accurate up to the (small) truncation
error. 
However, in order to optimize the basis states iteratively for
representing the target state(s) $\ket{\psi}$, new
information must be provided about $\ket{\psi}$,
i.e.\ it must be newly constructed using some unique criterion
(typically as the ground state of some Hamiltonian).
%This way, the reduced basis states  $\ket{\alpha}$ and  $\ket{\beta}$
%become at each step more and more optimized for representing
%$\ket{\psi}$.
Without such a criterion to ``sweep against'', the accuracy cannot
increase during sweeps, and the procedure would be pointless.
Merely transforming $\ket{\psi}$ in this way is therefore
of no use in standard DMRG, and is in fact never performed alone.
It is, however, the basis of the adaptive t-DMRG.

The adaptive t-DMRG algorithm relies on the Trotter decomposition of the time
evolution operator $U(t)$ of \eq{time_evo}, which is defined as follows:
Using the relation $U(t) = U(dt = t/M)^M$, the time evolution operator
is decomposed into $M$ time steps, where $M$ is a large number such
that the time interval $dt = t/M$ is small compared to the physical
time scales of the model.
Since the Hamilton operator of \eq{Hamilton} can be decomposed into a sum of
local terms $h_n$ that live only on sites $n$ and $n+1$, $U(dt)$ can 
then be approximated by an $n$-th order Trotter decomposition \cite{Suzuki1976},
e.g. to second order:
\be
\label{Trotter_2nd}
U(dt) =  \prod_{\stackrel{n}{\mathrm{even}}} \!\! U_n(\frac{dt}{2}) 
         \prod_{\stackrel{n}{\mathrm{odd}}} \!\! U_n(dt)  
         \prod_{\stackrel{n}{\mathrm{even}}} \!\! U_n(\frac{dt}{2})
+ O(dt^{3}).
\ee
The $U_n(dt)$ are the infinitesimal time-evolution operators $exp (-i h_n dt)$ 
on the bonds $n$ (even or odd). The ordering \emph{within} the even and odd
products does not matter, because ``even'' and ``odd'' operators commute
among themselves.

\Eq{Trotter_2nd} allows to decompose the time-evolution operator $U(t)$
into many local operators $U_n$ that live
on sites $n$ and $n+1$.
The adaptive time-dependent DMRG now allows to apply the 
operators $U_n$ successively to some state $\Psi$. 
Each operator $U_n$
is applied exactly during sweeps in the DMRG step 
with $n$ being the active site, i.e. where
sites $n$ and $n+1$ are represented without truncation (cf.\ \Eq{time_DMRG_state}):
This way, the basis states chosen to represent optimally the state
before $U_n$ is applied,
\be
\ket{\psi} = \sum_{\alpha \sigma \tau \beta} \psi_{\alpha \sigma \tau \beta} 
\ket{\alpha} \ket{\sigma}_n \ket{\tau}_{n+1} \ket{\beta}
\ee
are equally well suited for representing the state
\be
\label{state_after}
U_n \ket{\psi} = 
\sum_{\stackrel{\alpha \sigma \tau \beta}{\sigma' \tau'}}
(U_n)_{\sigma \tau,\sigma' \tau'} 
\psi_{\alpha \sigma' \tau' \beta} 
\ket{\alpha} \ket{\sigma}_n \ket{\tau}_{n+1} \ket{\beta}
\ee
without any additional error, because $U_n$ only acts on the part of
the Hilbert space (spanned by the vectors $\ket{\sigma}_n
\ket{\tau}_{n+1}$) 
that is exactly represented. 

To continue the sweep, a DMRG truncation
is carried out with $U_n\ket{\psi}$ being the target state instead of
$\ket{\psi}$. The key observation is that the new truncated basis is 
optimally adapted to
$U_n\ket{\psi}$ and different from the one that would have been chosen for
$\ket{\psi}$. 
In contrast to the conventional static DMRG \cite{CazalillaMarston02},
the optimally represented Hilbert space hence \emph{follows the
time-evolution of the state $\ket{\psi(t)}$}.

Then basis transformations to the left or right are performed,
until the next part of \eq{Trotter_2nd} can be applied. 
We thus apply the full operator of \eq{Trotter_2nd} by 
sweeping the active site $n$ through the system. 
The price to be paid is that a truncation error 
is introduced at each iteration step of the sweep 
as is known from static DMRG. 

To start time-dependent DMRG, some initial state has to
be prepared. There is no unique recipe, the most effective one
depending on the desired initial state. 
The procedure we adopt for our initial state  
$\ketini $ 
is to calculate it as the ground
state of a suitably chosen Hamiltonian $H_\mathrm{ini}$
(which does in principle not
have to have any physical significance).
Such a choice is
$H_\mathrm{ini} = \sum_n B_n S^z_n$, with $B_n <0$ for $n$ on the left,
$B_n >0$ for $n$ on the right half of the chain.
In this case, a physical picture for $H_\mathrm{ini}$ does exist; it
corresponds to switching on a magnetic field that aligns the spins and
that is strong enough for all interactions in \Eq{Hamilton} to be negligible.

\section{\label{time_error} Accuracy of the adaptive time-dependent DMRG}

As so far no quantitative analysis of the accuracy of the
adaptive t-DMRG has been given in the literature, 
we provide a detailed error analysis for the time evolution of
the initial state $\ketini$ in a spin-1/2 quantum XX chain,
i.e. $J_z = \delta = 0$. This system is an excellent benchmark 
for the adaptive t-DMRG due to its exact solution \cite{AntalSchutz99} that
can be compared to the DMRG results. The exact solution reveals
a nontrivial behaviour with a complicated substructure in the magnetization
profile. From a DMRG point of view this Hamiltonian is not too specific in
the sense that the experience from static DMRG suggests a relatively
weak truncation error dependence on $J^z$.

\subsection{Possible errors}
Two main sources of error occur in the adaptive t-DMRG:
\\ (i) The {\em Trotter error} due to the Trotter decomposition. For
an $n$th-order Trotter decomposition
\cite{Suzuki1976}, the error made in one time step 
$dt$ is of order $dt^{n+1}$. To reach a given time $t$ one has to perform 
$t/dt$ time-steps, such that in the worst case the error grows linearly in time $t$ and
the resulting error is of order $(dt)^n t$. 
In our setup of the
Trotter decomposition, the error scales linearly with system size
$L$, and overall it is of order $(dt)^n L t$ for the times of
interest. (Eventually, the error must saturate at a finite value, 
as measured quantities are typically bounded.)
The linear $L$ dependence of the error is expected for
generic initial states. For the particular choice of $\ketini$ of this
paper, however, many of the $O(L)$ contributions to the Trotter error
vanish, as many of the sites exhibit no dynamics at all for short times.
-- For the calculations presented in this paper, we have chosen
$n=2$, but our observations should be generic. 
%Note that at given orders
%of Trotter decompositions, different choices lead to possibly much larger 
%prefactors (cf. \olcite{WhiteFeiguin04,Suzuki1976}).
\\
(ii) The DMRG {\em truncation error} due to the representation of the
time-evolving quantum state in reduced (albeit ``optimally'' chosen) 
Hilbert spaces and to the repeated transformations between different
truncated basis sets.
While the truncation error $\epsilon$ that sets the scale of
the error of the wave function and operators is typically very small, 
here it will strongly accumulate
as $O(Lt/dt)$ truncations are carried out up to time $t$. This is because
the truncated DMRG wave function has norm less than one and is renormalized
at each truncation by a factor of $(1-\epsilon)^{-1}>1$. Truncation errors
should therefore accumulate roughly exponentially with an exponent
of $\epsilon Lt/dt$, such that 
eventually the adaptive t-DMRG will break down at too long times.
The error measure we use here saturates at $O(1)$ and sets a
limit on the exponential growth; also, partial compensations of errors 
in observables may slow down the error growth.
The accumulated truncation error should decrease considerably with an 
increasing number of kept DMRG states $m$. For a fixed time $t$, 
it should decrease as the Trotter time step $dt$ is increased, as the number 
of truncations decreases with the number of time steps $t/dt$.

At this point, it is worthwhile to mention that our subsequent error analysis
should also be pertinent to the very closely related time-evolution
algorithm introduced by Verstraete {\em et al.}\cite{Verstraete04}, which 
differs from ours for the present purpose in one major point: In our
algorithm a basis truncation is performed after each  
{\em local} application of $U_n$. In their algorithm truncations are 
performed after all local time-evolutions have been carried out, 
i.e.\ after a {\em global} time-evolution using $U = \prod_n U_n$. 
In our iterative procedure, 
the wave function after such a full time evolution
is not guaranteed to be the \emph{globally} optimal state
representing the time-evolved state. However, 
for small $dt$ the state update via 
the operators $U_n$ is likely to be small, we expect the global optimum to be
rather well approximated using the present algorithm, as 
seems to be borne out by direct comparisons between both approaches
\cite{Verstraeteprivcom}. Errors should therefore exhibit very similar
behaviour.

We remind that no error is encountered in the application of
the local time evolution operator $U_n$ to the state $\ket{\psi}$, as
is discussed after \Eq{state_after}.

\subsection{Error analysis for the XX-model}

In this section, we analyze the errors from the adaptive t-DMRG in the time
evolution of the XX-model by comparing it to the exact solution
\cite{AntalSchutz99}, with the ultimate goal of finding optimal
DMRG control parameters to minimize the errors.

We use two main measures for the error: 
\\ (i) As a measure for the overall error we consider the {\em magnetization
deviation} the maximum deviation of the local magnetization found by DMRG from
the exact result, 
\be
\mathrm{err}(t) = 
\mathrm{max}_n |\langle S^z_{n,\mathrm{DMRG}}(t)\rangle -
\langle S^z_{n,\mathrm{exact}}(t)\rangle |.
\ee
In the present study, the maximum was typically found close to the center of 
the chain.
\\ (ii) As a measure which excludes the Trotter error we use the 
{\em forth-back deviation} $FB(t)$, 
which we define as the deviation between the initial state 
$\ketini$ and the state $\ket{fb(t)} = U(-t)U(t) \ketini$,
i.e.\ the state obtained by evolving $\ketini$ to some time $t$ and
then back to $t=0$ again. 
If we Trotter-decompose the time evolution operator $U(-t)$ into
odd and even bonds in the reverse order of the decomposition of $U(t)$,
the identity $U(-t) = U(t)^{-1}$ holds without any Trotter error, and
the forth-back deviation has the appealing property to capture the
truncation error only.
In contrast to the magnetization deviation, the forth-back error does
not rely on the existence of an exact solution.

As our DMRG setup does not allow easy access to the fidelity
$|\bra{\init}fb(t) \rangle|$, we define the forth-back deviation 
to be the $L_2$
measure for the difference of the magnetization profiles of
$\ketini$ and $\ket{fb(t)}$,
\be
FB(t) = \left( \sum_n \left( \bra{\mathrm{ini}} S^z_n \ketini - \bra{fb(t)}
S^z_n \ket{fb(t)} \right)^2 \right)^{1/2}
\!\!\!\!\!\!\!.
\ee

\begin{figure}
\epsfig{file=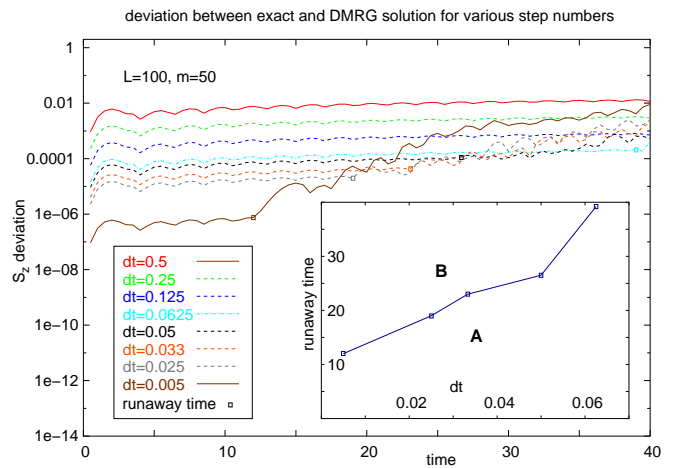, width=1\linewidth}
\caption[Magnetization deviation as a function of time for different
Trotter time steps $dt$.]{ \label{eval_m50_dt}
Magnetization deviation as a function of time for different Trotter time steps
$dt$ and for $m=50$ DMRG states. At small times (region A in the
inset), the deviation is dominated by  
the linearly growing Trotter error for small times. At later times
(region B in the inset), much faster,
non-linear growth of the deviation sets in at some well-defined runaway-time
$t_R$. As shown in the inset, $t_R$ increases with increasing $dt$.
}
\end{figure}

\begin{figure}
\epsfig{file=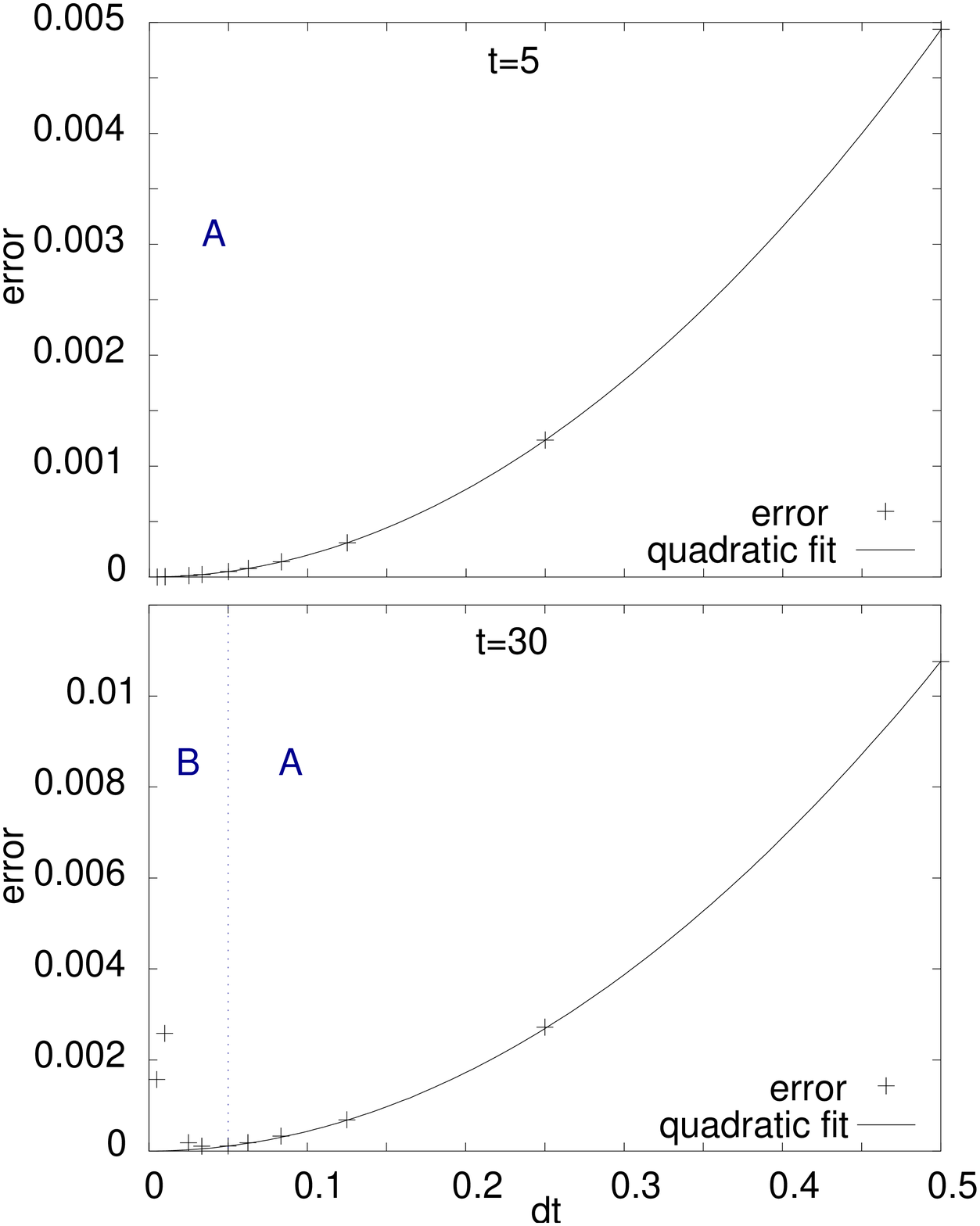, width=.9\linewidth}
 \caption[Magnetization deviation as a function of  Trotter time step
 $dt$ at times $t=5$ (upper figure) and $t=30$.]{ \label{Trotter_error}
Magnetization deviation as a function of  Trotter time step $dt$ 
(system size $L=100$, $m=50$ DMRG states) at times $t=5$ (upper figure) and 
$t=30$ (lower figure). For $t=5$, the magnetization deviation is
quadratic in $dt$ as expected from the Trotter error. For $t=30$, at
small $dt$ the magnetization deviation is no longer quadratic in $dt$ and
larger than the Trotter error would suggests. This is a signal of the
contribution of the accumulated truncation error.
}
\end{figure}

In order to control Trotter and truncation error, two DMRG control parameters
are available, the number of DMRG states $m$ and the Trotter time step $dt$.

To study the effect of varying $dt$, consider the 
{\em magnetization deviation} as shown in \Fig{eval_m50_dt}. Two
main observations can be made. At small times (regime A), 
the magnetization deviation
decreases with $dt$ and is linear in $t$ as expected from the Trotter error. 
Indeed, as shown
in the upper part of \Fig{Trotter_error}, the magnetization deviation
depends quadratically on $dt$ for fixed $t$, 
and the Trotter error dominates over the truncation error. 
At large times (regime B), the magnetization deviation is no longer
linear in $t$, but grows almost exponentially, and also does no longer
show simple monotonic behaviour in $dt$:
The magnetization deviation in this regime is obviously no longer
dominated by the Trotter error, but by the accumulated truncation error.

The two regimes A and B are very clearly separated by some {\em runaway time}
$t_R$, with regime A for $t<t_R$ and regime B for $t>t_R$ (a precise procedure
for its determination will be outlined below). 
The runaway time $t_R$ increases when $dt$ is increased:
Because the total number of Trotter time steps $t/dt$ is decreased,
the accumulated truncation error decreases, and the Trotter error
increases, hence the competing two errors break even later.

This $dt$-dependence of $t_R$ is also seen in the 
lower part of \Fig{Trotter_error}, where the $dt$ dependence of the 
magnetization deviation is plotted at some larger time ($t=30$) than in 
the upper part. $t=30$ is larger than the runaway time (i.e.\ in regime B) 
for $dt \leq 0.05$, in regime A otherwise. 
We see indeed for $dt > 0.05$
  (region A) the familiar quadratic 
Trotter error dependence. For small $dt \leq 0.05$ (region B), the 
deviation is dominated by the accumulated truncation error that increases
as $dt$ decreases. This is reflected in the growth of the
magnetization deviation as $dt$ is \emph{de}creased.

\begin{figure}
\epsfig{file=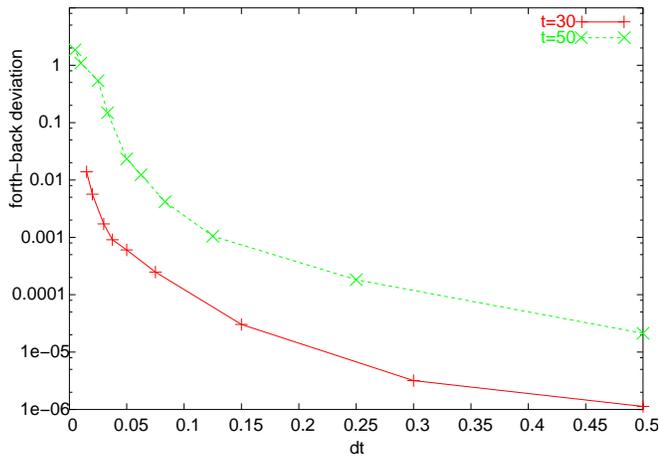, width=1\linewidth}
\caption[The forth-back error $FB(t)$ for $t=30$ and $t=50$, as
function of $dt$.]{ 
\label{backforth_dt_fig} The forth-back error $FB(t)$ for $t=30$ and $t=50$, as function of
$dt$. Here, $L=100$, $m=50$.
}
\end{figure}

The almost exponential growth of the truncation error with the number
of Trotter steps can also be seen from 
the forth-back deviation that is not susceptible to the Trotter error. 
In \Fig{backforth_dt_fig}, we show the forth-back deviation $FB(t)$ for
$t=30$ and $t=50$ as a function of the Trotter time step $dt$. $FB(t)$ 
increases as a consequence of the stronger accumulation of the truncation error
with decreasing Trotter step size $dt$ and hence an increasing number of
steps $t/dt$.

\begin{figure}
\epsfig{file=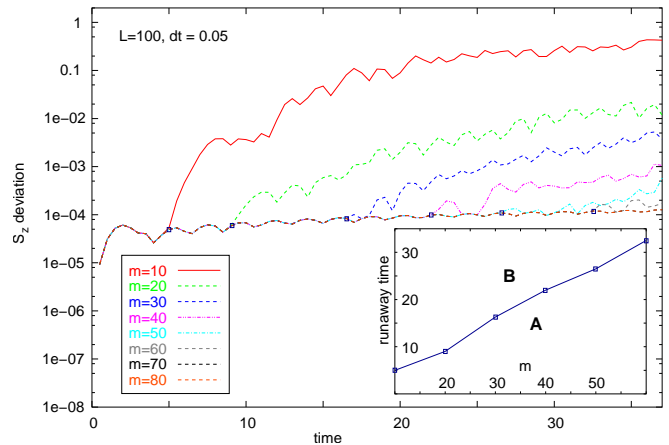, width=1\linewidth}
 \caption[Magnetization deviation $\Delta M(t)$ as a function of time for
different numbers $m$ of DMRG states.]{ \label{eval_T50_m}
Magnetization deviation $\Delta M(t)$ as a function of time for
different numbers  
$m$ of DMRG states. The Trotter time interval is fixed at $dt = 0.05$. 
Again, two regimes can be distinguished: 
For early times, for which the Trotter error dominates, the error is slowly  
growing (essentially linearly) and independent of $m$ (regime A); 
for later times, the error is entirely given by the truncation error,
which is $m$-dependent and growing fast 
(almost exponential up to some saturation; regime B).
The transition between the two regimes occurs at a well-defined
``runaway time'' $t_R$ (small squares). The inset shows a monotonic, roughly
linear dependence of $t_R$ on $m$.}
\end{figure}

Let us now consider the dependence of the magnetization deviation $\mathrm{err}(t)$
on the second control parameter, the number $m$ of DMRG states.
In \Fig{eval_T50_m}, $\mathrm{err}(t)$ is plotted for a fixed Trotter time 
step $dt=0.05$ and different values of $m$. 
In agreement with our previous
observations, some $m$-dependent ``runaway time'' $t_R$, separates two
regimes: for $t<t_R$ (regime A), the deviation grows essentially linearly
in time and is independent of $m$, for $t>t_R$ (regime B), it suddenly
starts to grow more rapidly than any power-law.
The onset of a significant $m$-dependence has indeed been our
operational definition of $t_R$ in \Fig{eval_m50_dt} and \ref{eval_T50_m}.
In the inset of \Fig{eval_T50_m}, $t_R$ 
is seen to increase roughly linearly with growing $m$. As $m\rightarrow\infty$
corresponds to the complete absence of the truncation error, the
$m$-independent bottom
curve of \Fig{eval_T50_m} is a measure for the deviation due to the Trotter 
error alone and the runaway time can be read off very precisely as the moment
in time when the truncation error starts to dominate. 

\begin{figure}
\epsfig{file=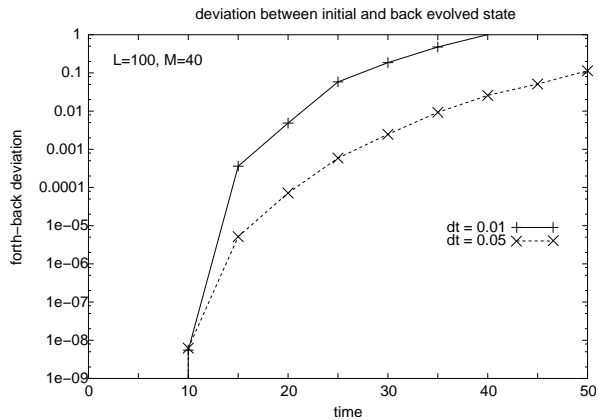, width=.9\linewidth}
\caption[The forth-back error as function of $t$.]{ \label{backforth_t_fig}
The forth-back error $FB(t)$ for $L=100$, $m=40$, $dt=0.01$ and $dt = 0.05$, as
function of $t$.
}
\end{figure}

\begin{figure}
\epsfig{file=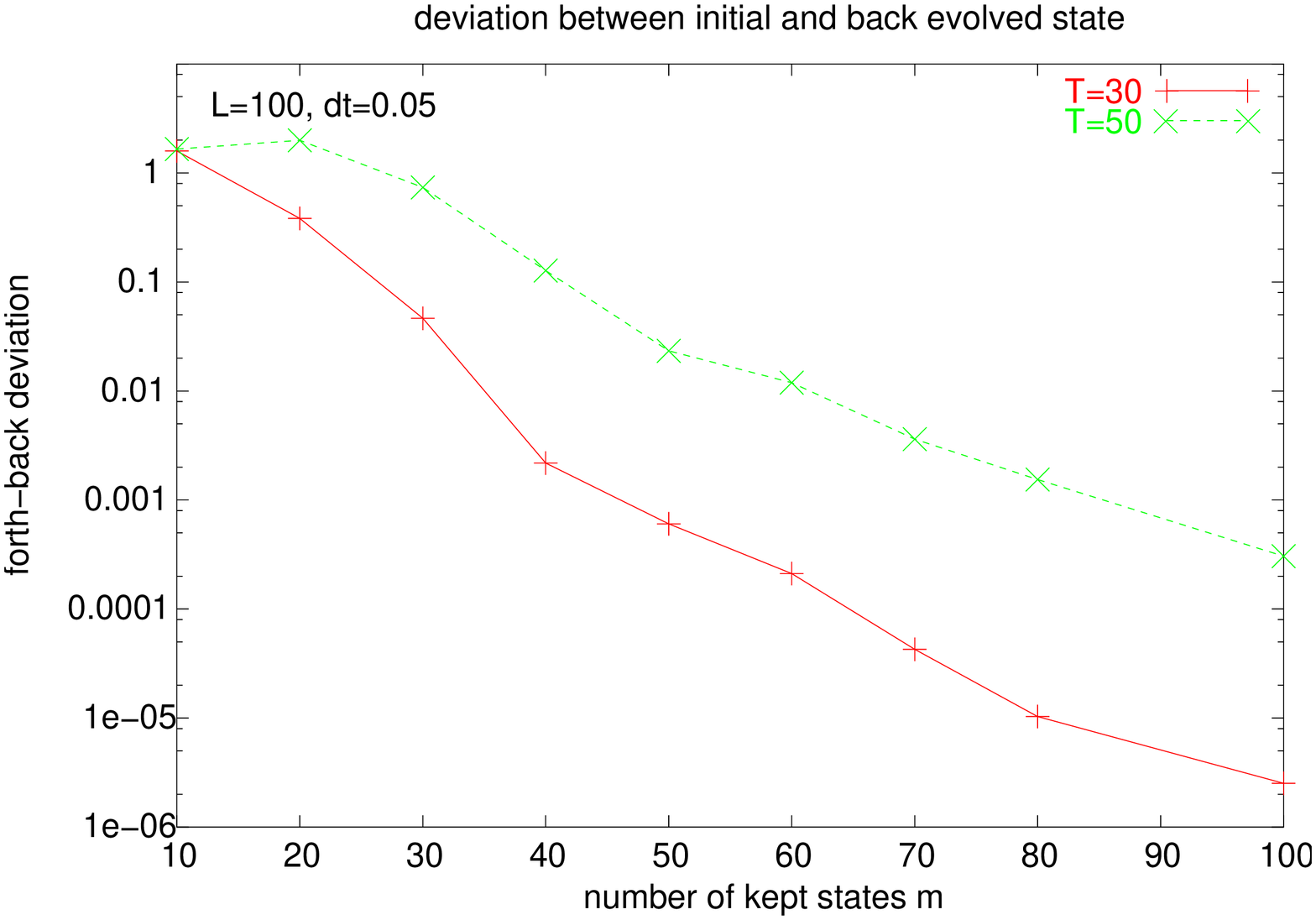, width=.9\linewidth}
\caption[The forth-back error as function of $m$.]{ \label{backforth_m_fig}
The forth-back error $FB(t)$ for $t=50$ and $t=30$, as function of
$m$.
Here, $L=100$, $dt = 0.05$.
}
\end{figure}

That the crossover from a dominating Trotter error at short times and a
dominating truncation error at long times is so sharp may seem surprising
at first, but can be explained easily by observing that the Trotter error
grows only linearly in time, but the accumulated truncation error grows 
almost exponentially in time.
The latter fact is shown in \Fig{backforth_t_fig}, where the forth-back 
deviation $FB(t)$ is plotted as a function of $t$ for some fixed $m$.
Here, we find that the effects of the truncation error are 
below machine precision for $t<10$ and then grow almost exponentially in time 
up to some saturation. 

By comparison, consider \Fig{backforth_m_fig}, where $FB(t)$ is plotted as 
a function of $m$, for $t=30$ and $t=50$. An approximately exponential increase 
of the accuracy of the method with growing $m$ is observed for a fixed time.
Our numerical results that indicate a roughly linear time-dependence of
$t_R$ on $m$ (inset of \Fig{eval_T50_m}) are the consequence of some
balancing of very fast growth of precision with $m$ and decay of precision
with $t$. 

Before concluding this section, let us briefly consider a number of
other possible effects that might affect $t_R$. One might alternatively
conceive that the well-defined runaway-time $t_R$ results from a sudden failure
(of stochastic or of fundamental nature) of the truncation algorithm
to capture one important basis state.
It can be refuted on the basis of
\Fig{backforth_dt_fig}, \Fig{backforth_t_fig} and \Fig{backforth_m_fig}: 
Such an error should
manifest itself as a pronounced step in $FB(t)$, depending on the time
evolution having gone past $t_R$ or not. Such a step is, however, not observed.

\begin{figure}
\epsfig{file=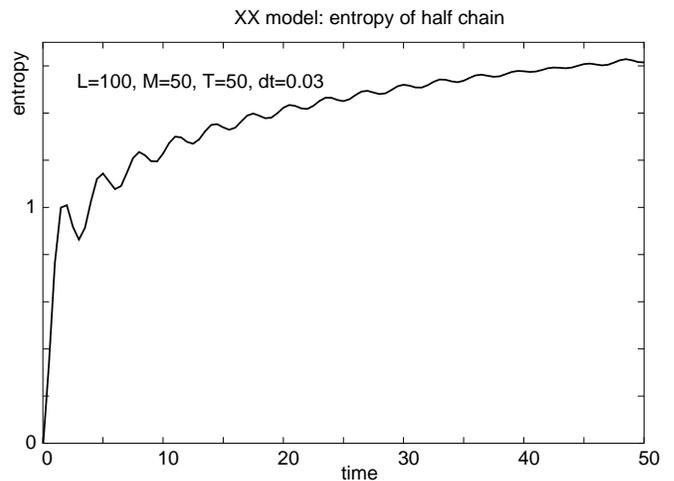, width=1\linewidth}
 \caption[Entanglement entropy of the left (or, equivalently, the right) half of the chain.]{ \label{entropy}
Entanglement entropy $S_e$ from \Eq{entropy_Eq} between 
the left and the right half of the chain as function of time.
}
\end{figure}

$t_R$ might also be thought to reflect a fundamental DMRG limit,
namely a growth of the entanglement within the time-evolved state
which the limited number of DMRG states $m$ is not able to capture
adequately at $t>t_R$. This scenario can be excluded by observing the 
strong dependence of
$t_R$ on the number of time steps, which this scenario cannot explain.
Indeed, a study of the entanglement entropy between the left and the
right half of the chain
\be
\label{entropy_Eq}
S_e(t) = \mathrm{Tr} \hat{\rho} \mathrm{log}_2 \hat{\rho},
\ee 
$\hat{\rho}$ being the reduced density matrix of the left (or
equivalently the right) half of the chain,
confirms this view:
As shown in \Fig{entropy}, $S_e(t)$ is only mildly
growing with time and is well below the maximum entanglement entropy
$S_\mathrm{max} \sim \mathrm{log}_2 m$ that the DMRG can reproduce.

Therefore we conclude that the error at short times is dominated by the
Trotter error, which is independent of $m$ and approximately growing
linearly with time.  
At some runaway time, we observe a sharp crossover to a regime in which
the $m$-dependent and almost expontially growing truncation error is
dominating.
This crossover is sharp due to drastically different growth of the
two types of errors.
The runaway time thus indicates an imminent
breakdown of the method and is a good, albeit very conservative measure of
available simulation times. We expect the above error analysis for the 
adaptive t-DMRG to be generic for other models. The truncation error
will remain also in approaches that dispose of the Trotter error; maximally
reachable simulation times should therefore be roughly the same or somewhat
shorter if other approximations enhance the truncation error.

\subsection{Optimal choice of DMRG parameters}

\begin{figure}
\epsfig{file=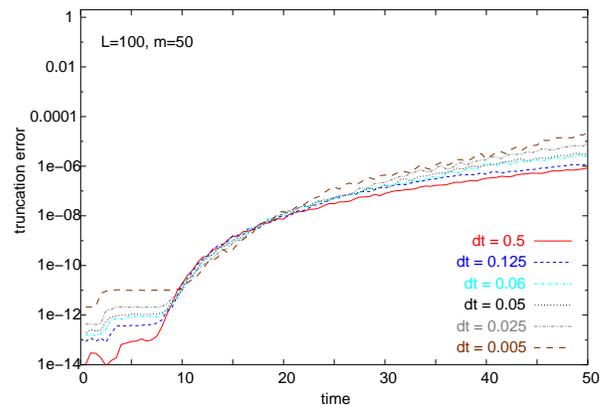, width=.9\linewidth}
 \caption[The lost weight in the density matrix truncation, summed over time intervals $\Delta t = 0.1$.]{ \label{eval23_m50_dt}
The lost weight in the density matrix truncation, 
%(upper fig.) and the norm reduction of the truncated state (lower
%fig.), 
summed over time intervals $\Delta t = 0.1$, is shown for the same
parameters as in \Fig{eval_m50_dt}.
%Both curves are highly correlated. 
A comparison with
\Fig{eval_m50_dt} reveals, however, that both values are not useful
criteria for the DMRG truncation error and are in particular not
suited to reveal the runaway time $t_R$.
%TODO: Make nicer, remove some spaghetti.\new{comment on accuracy, down to where to be plotted?}
}
\end{figure}

How can the overall error -- which we found to be a delicate balance between the
Trotter and the accumulated truncation error -- 
be minimized and the important runaway time be found in practice? 
From the above scenario it should be expected that the truncated density 
matrix weight at each step does not behave differently before or after the 
runaway time and hence is no immediately useful indicator to identify the
runaway time. This can
in fact be seen from \Fig{eval23_m50_dt}, where the truncated weight is
shown for the same parameters as in \Fig{eval_m50_dt}. Also, it is not
obvious to extract a precise relationship between the truncation errors at each
DMRG truncation and the accumulated errors. 
Instead, a precise convergence analysis in $m$ or $dt$ seems to be
more telling and easily feasible.

Of course, it is desirable to choose the number of kept states $m$
as large as possible within the constraints regarding the available
computer resources. 
This choice having been made, 
the runaway time $t_R$ is determined for different Trotter time
steps $dt$ by comparing different values of $m$ as in
\Fig{eval_T50_m}. 
Only two slightly different values of $m$ are sufficient for that purpose. 
Now the Trotter time step $dt$ is chosen such that the desired time $t$
is just below $t_R$. 
This way, the optimal balance between the Trotter error and the
truncation error is found, which corresponds in the lower part of
\Fig{Trotter_error} to the minimum of $\mathrm{err}(t)$ on the border
between regime A and B:
The total error would increase at larger $dt$ due to the Trotter
error, and at smaller $dt$ due to the truncation error.

Thus,  it is a good practice to choose for small times rather small values of
$dt$ in order to minimize the Trotter error; for large times, it makes
sense to choose a somewhat coarser time interval, in order to push the
runaway time to as large values as possible.

In terms of numbers of time steps, we conclude from \Fig{eval_m50_dt}
that for the present model and our parameters ($L=100-200$), the 
adaptive time-dependent DMRG seems to be able to perform about 1000-5000 time
steps reliably even for $m=50$, depending on the desired level of
accuracy, corresponding to $O(100/J)$ in ``real'' time.
We note that this is a very small value of $m$ by DMRG standards, and
that using an optimized code, one should be able to increase $m$
by an order of magnitude, and hence access much longer times (by an
order of magnitude).

\section{\label{time_scaling} Long-time properties of the time-evolution}

In \olcite{AntalSchutz99, HunyadiSasvari04}, the time evolution of the
initial state $\ket{\init}$ on the XX chain  at
temperature $T=0$ was examined in the long-time limit using the exact
solution.
It was found that the magnetization $S_z(n,t)$ given in \Eq{Sz_nt}
can be described for long times  
in terms of a simple scaling function, 
$S_z(n,t) \approx \Phi\left( (n-n_c)/t \right)$, where $n_c$ is the
position of the chain center.
The scaling function is the solution of the partial differential equation
$\partial_t S_z + \partial_x j(S_z)=0$ with the magnetization
current  $j(S_z)= 1/\pi \cos{|\pi S_z|}$ which has been shown to describe
the macroscopic time evolution of the magnetization profile
\cite{AntalSchutz99}.
The characteristics, i.e.\ the lines of constant magnetization $S_z$,
have a slope $v=\sin{|\pi S_z|}$.

The magnetization profile $\Phi\left( (n-n_c)/t \right)$ has a
well-defined front at $(n-n_c)/t = \pm 1$,
i.e.\ is moving outwards ballistically with velocity $v=1$.
On top of this overall scaling form an additional step-like 
substructure arises, which was analysed in detail in 
\olcite{HunyadiSasvari04}. It was found that while
the step width broadens as $t^{1/3}$, the step height decreases as
$t^{- 1/3}$,
such that the integrated transported magnetization within each step 
remains constant at 1. It was suggested that each of these steps
corresponds to a localized flipped spin flowing outwards.

The XX model, however, has several very special properties:
It corresponds to a free-fermion model and is therefore exactly
solvable; it is critical; and its total current operator $J=\sum_n
j_n$ commutes with the Hamiltonian, $[J,H]=0$. 
One may ask whether the above 
findings are due to any of the particularities of the XX model or more
generic.

The adaptive t-DMRG allows us to study the long-time evolution of 
$\ket{\textrm{ini}}$ in different coupling regimes of \eq{Hamilton}. 
We chose two extensions of the XX model, namely a $S^zS^z$-
interaction, and dimerization.

\begin{figure}
\epsfig{file=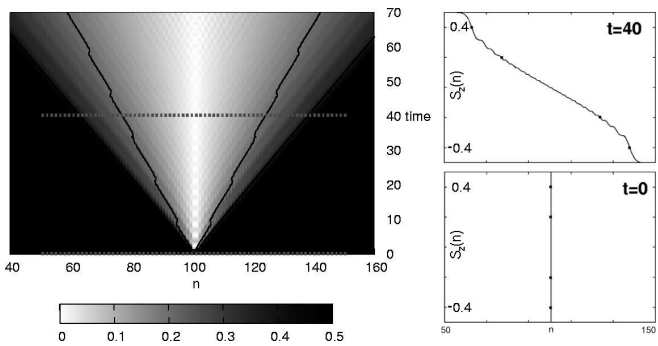, width=1\linewidth}
\caption[Time evolution of the absolute value of the 
local magnetization $|\langle S^z_n(t) \rangle|$ 
for the XX model as a density plot.]{ \label{Jz0_Fig}
Left: Time evolution of the absolute value of the 
local magnetization $|\langle S^z_n(t) \rangle|$ 
for the XX model as a density plot, where the local magnetization itself
is exactly antisymmetric with regard to the chain center. 
The lines of constant-magnetization $\langle S^z_n \rangle = \pm 0.2, \pm
0.4$ are shown as solid lines.
As an illustration, local magnetizations $\langle S^z_n(t) \rangle$ for 
the time slices $t=0$ and $t=40$ are shown
explicitly.
A step-like substructure can be seen for $t=40$ in perfect quantitative
agreement with the exact solution. Error bars are below visibility.
}
\end{figure}
\begin{figure*}
\epsfig{file=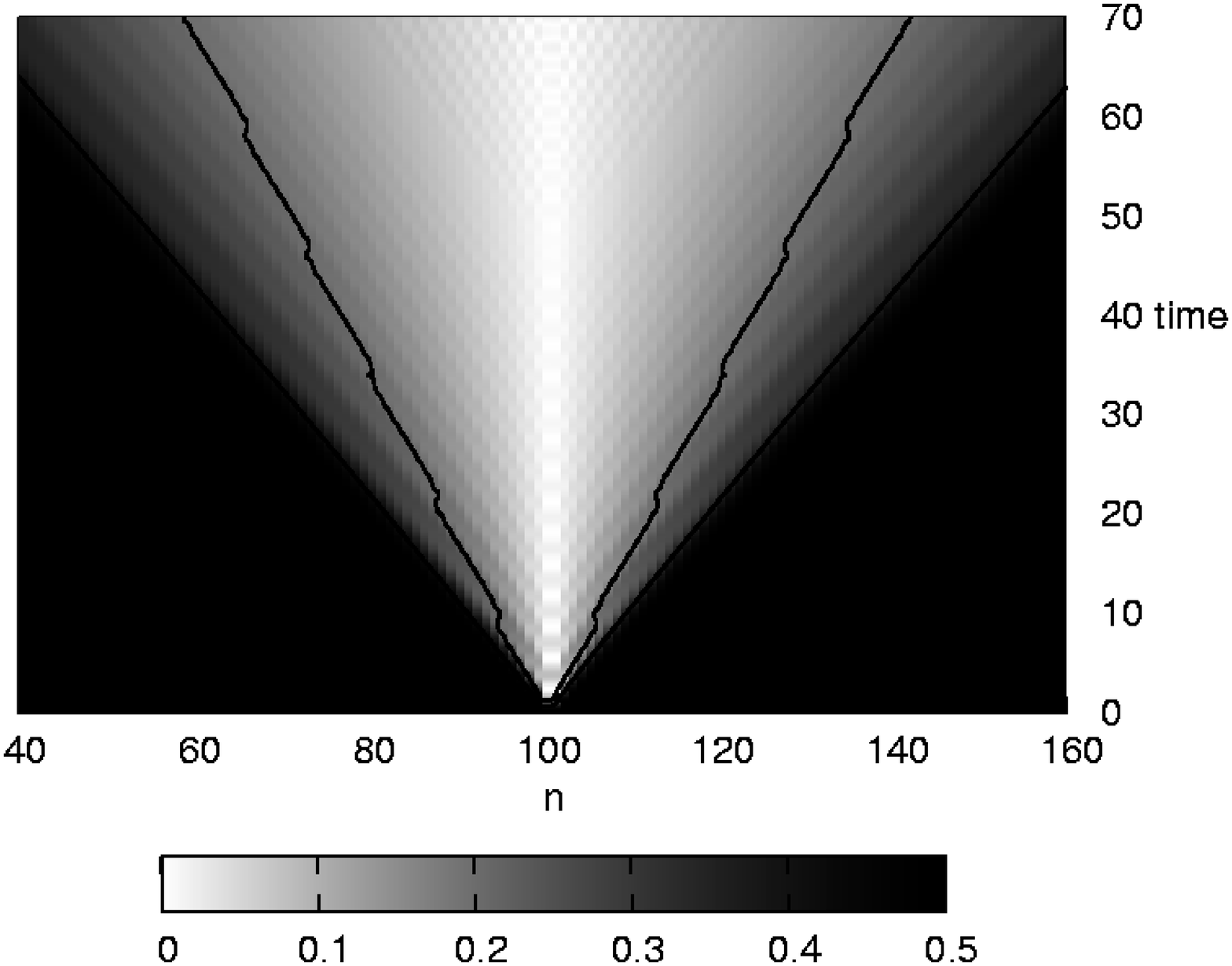, width=.32\linewidth}
\epsfig{file=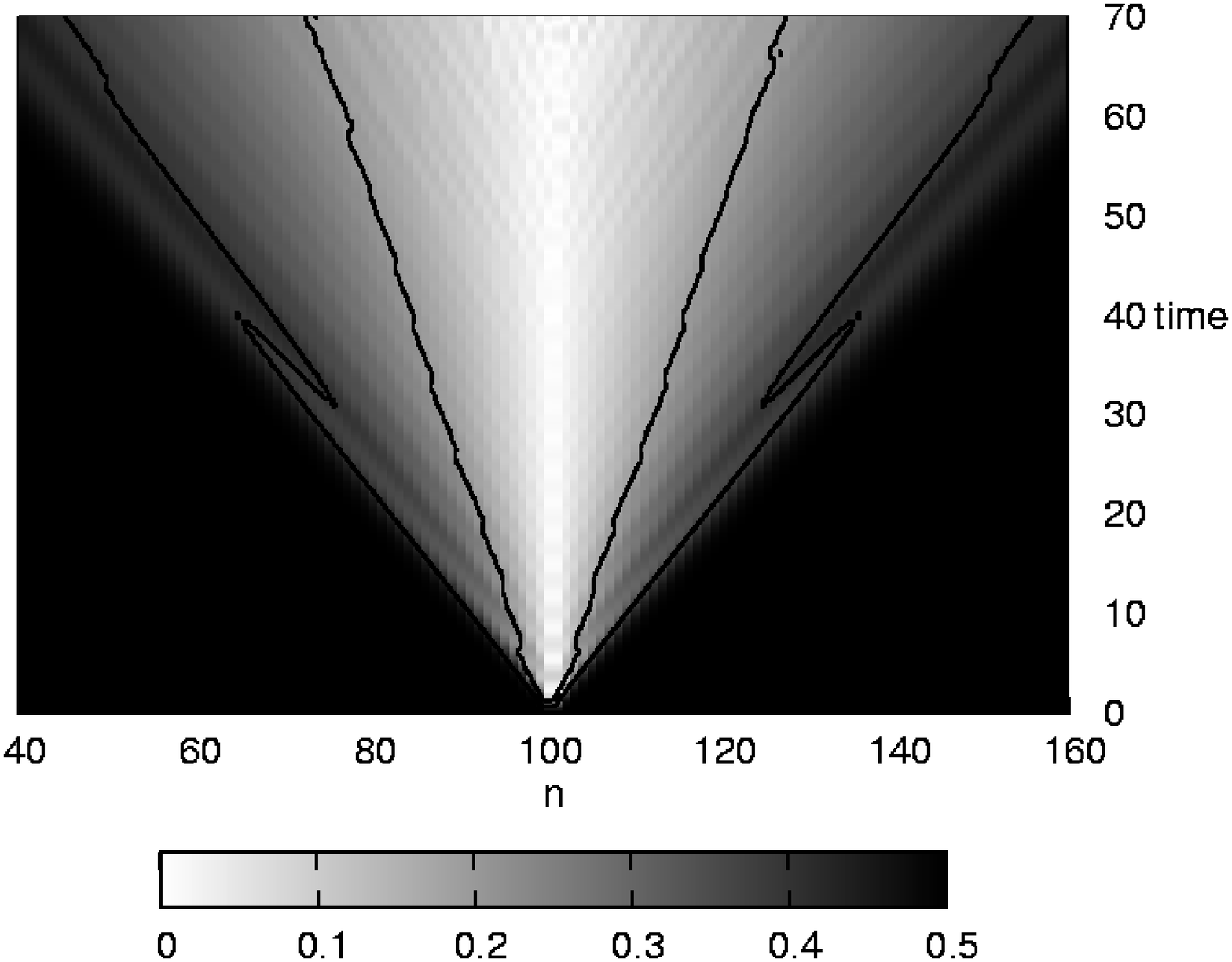,  width=.32\linewidth}
\epsfig{file=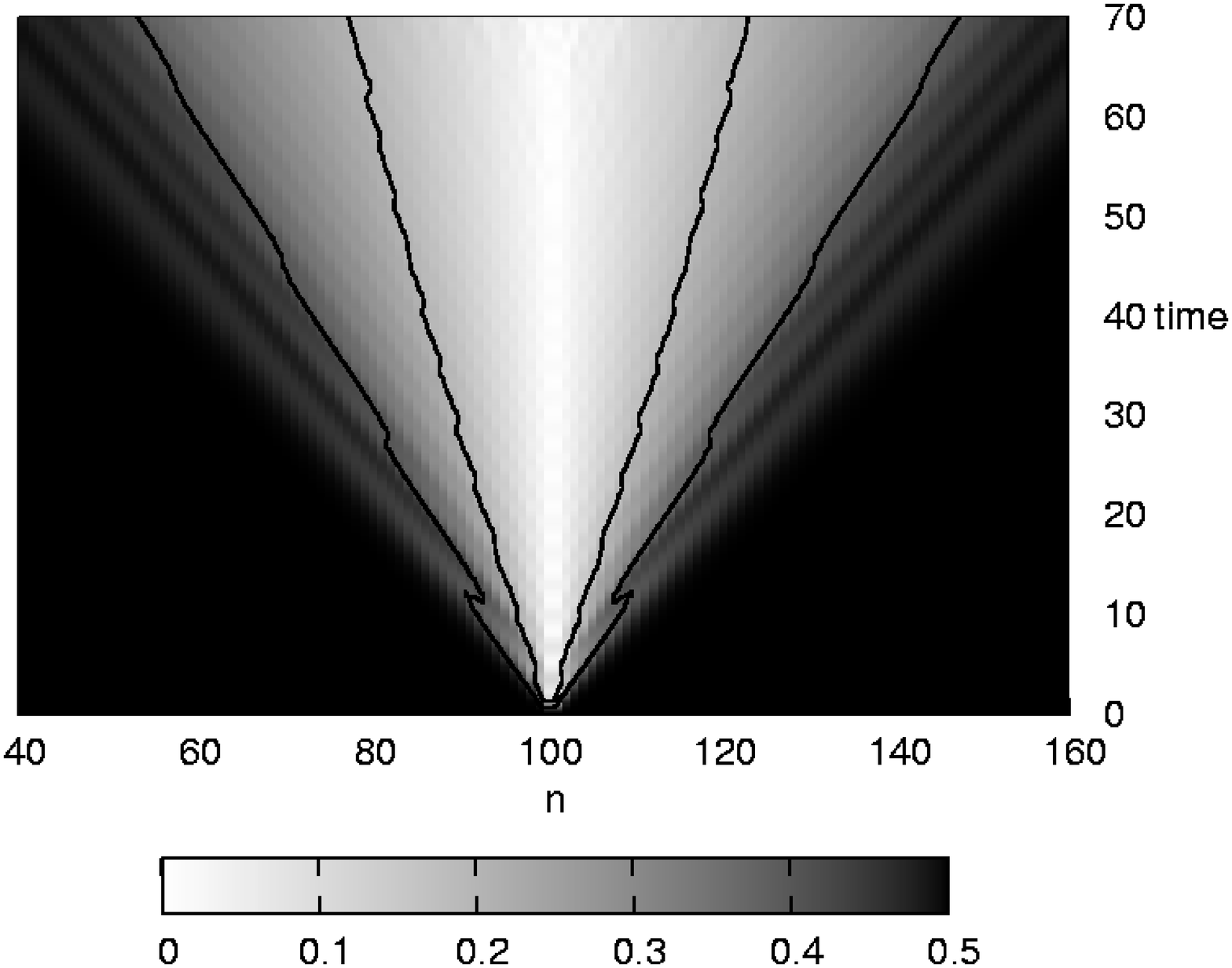,  width=.32\linewidth}
\epsfig{file=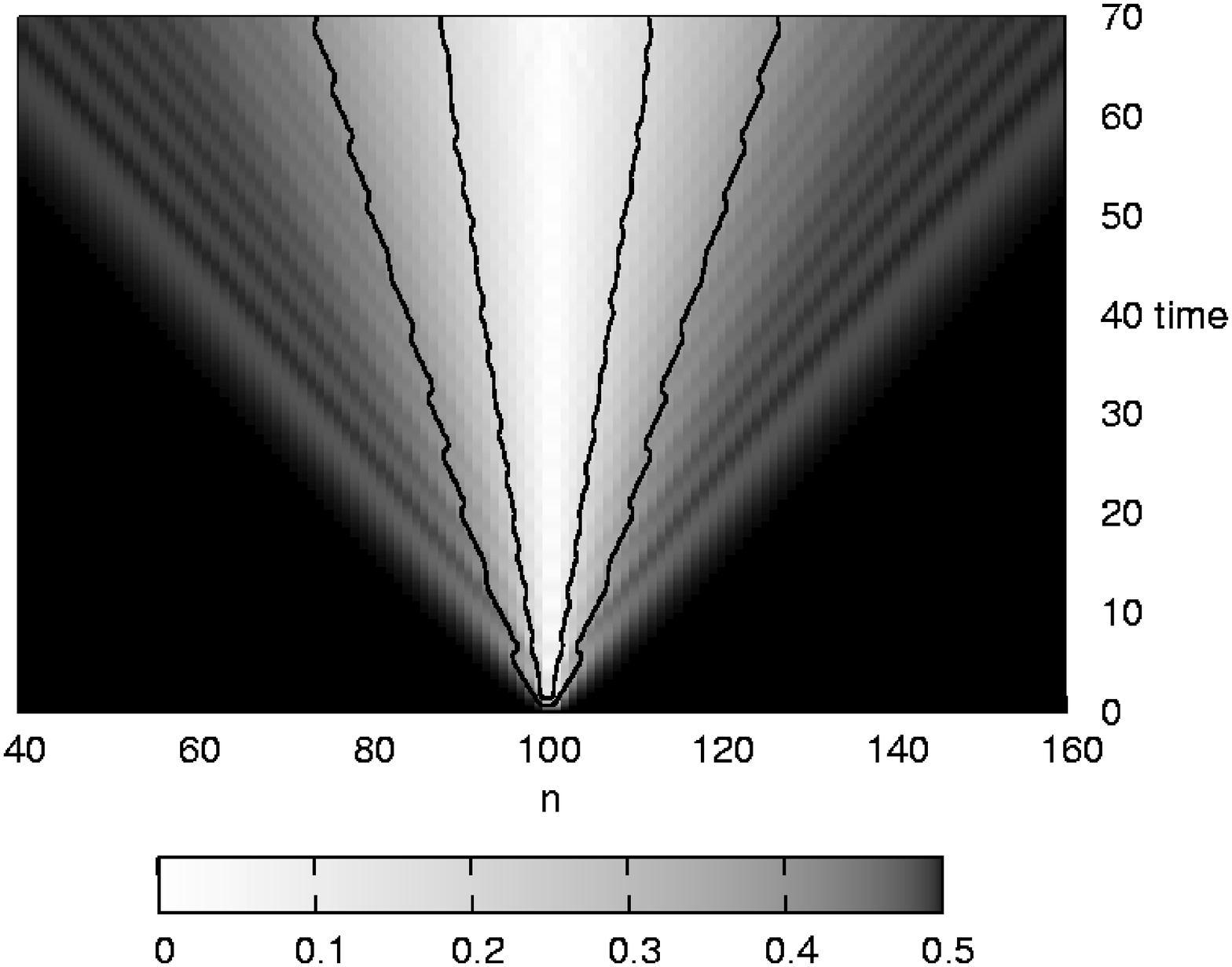,  width=.32\linewidth}
\epsfig{file=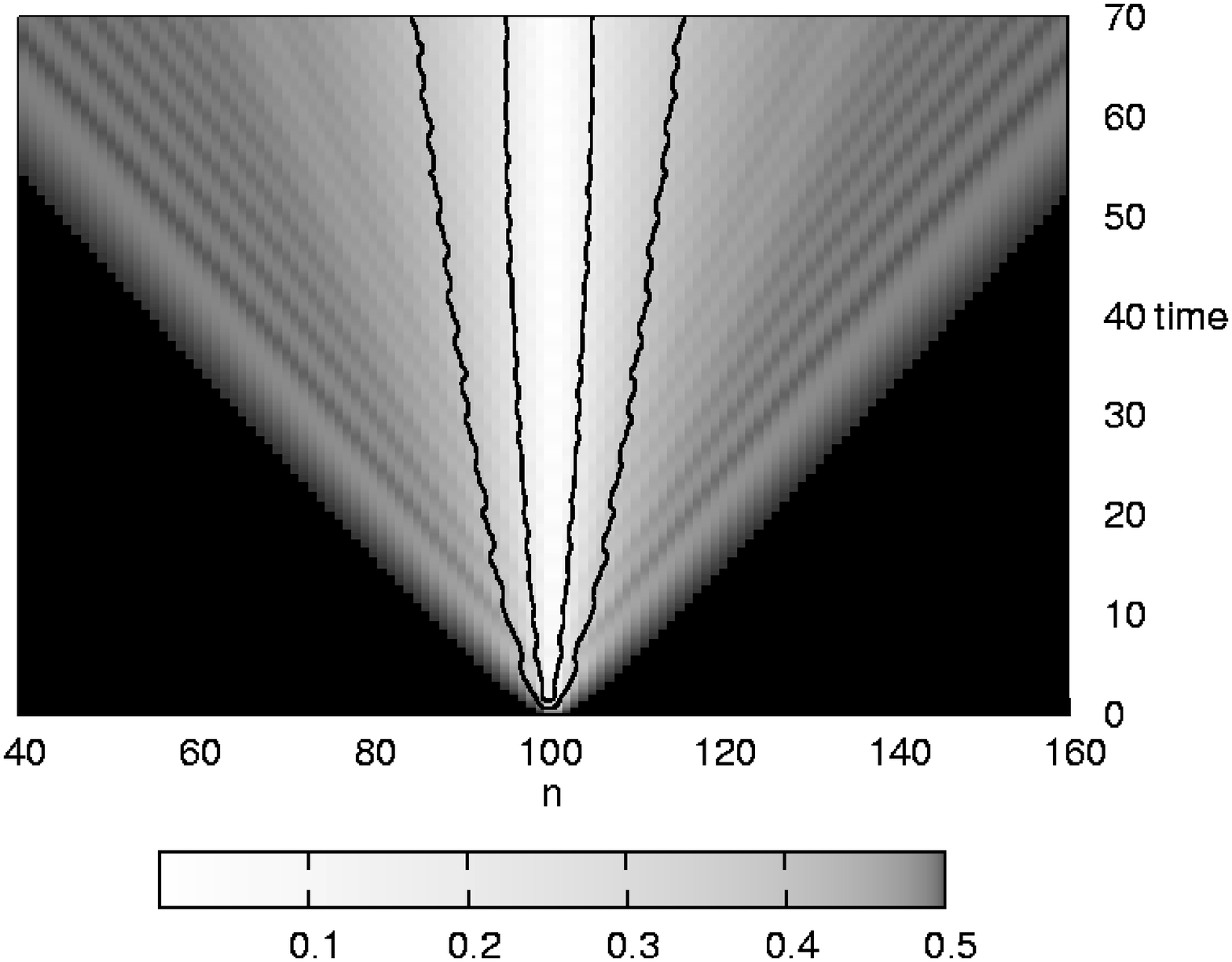,  width=.32\linewidth}
\epsfig{file=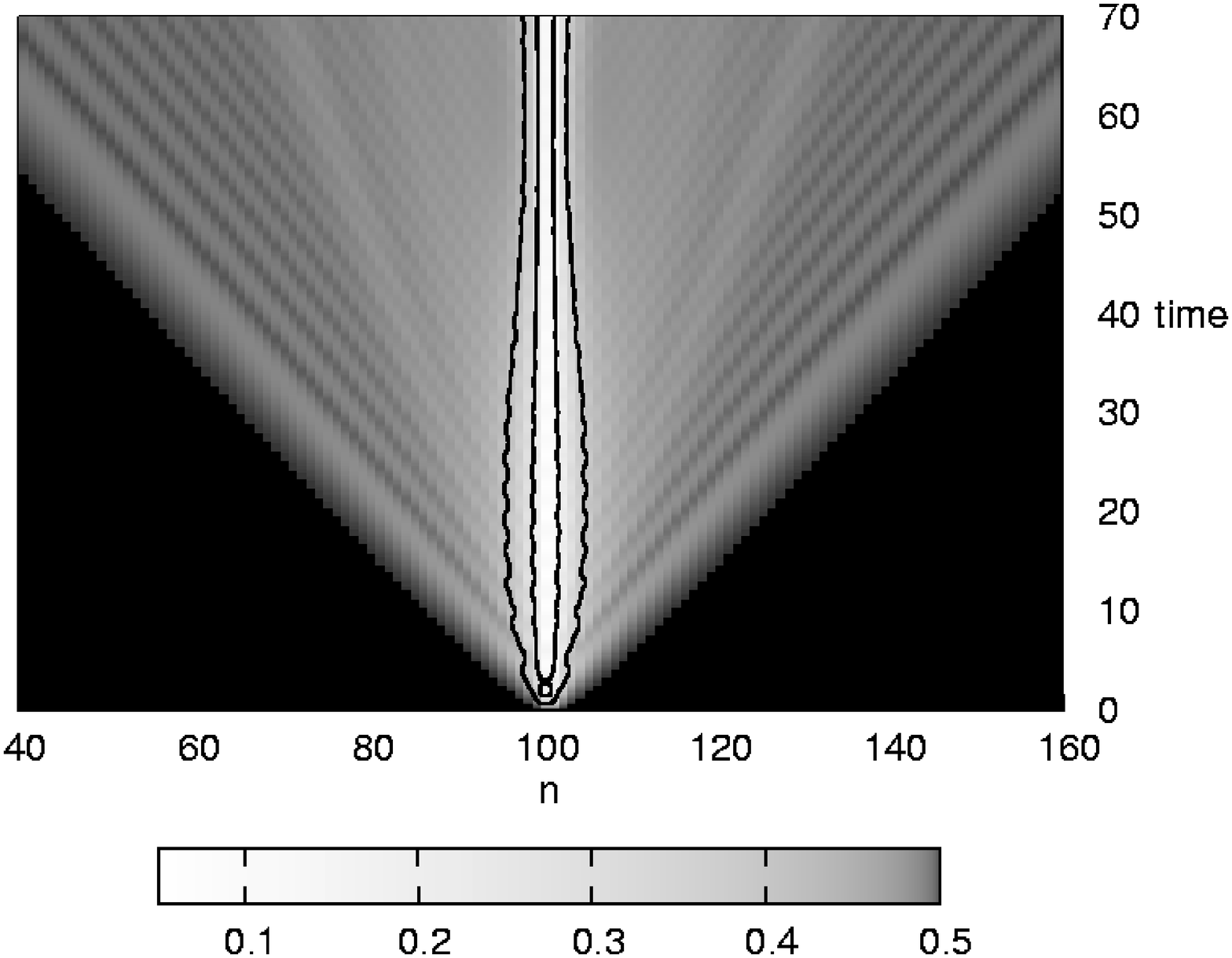,  width=.32\linewidth}
\caption[Density plots of the magnetization $|\langle S^z_n(t)
\rangle|$ for $J_z=$ 0, 0.3, 0.6, 0.9, 1.0, 1.1, and $\delta = 0$.]{ 
\label{Jz_Fig}
Density plots of the magnetization $|\langle S^z_n(t) \rangle|$ as in
\Fig{Jz0_Fig}, the 
values of $J_z$ being (from left to right, top to bottom) 0, 0.3, 0.6,
0.9, 1.0, 1.1, and $\delta = 0$.
For better visibility of the profile, the grayscale mapping of
$|\langle S^z_n(t)\rangle |$ 
was chosen differently in each plot as indicated by the
legends.
Solid lines: lines of constant magnetization $\langle S^z_n \rangle =
\pm 0.2, \pm 0.4$; these allow for a direct comparison of the
magnetization transport between different $J_z$. The ray-like structure
indicates the ``carriers''.
%Fat dots: The front position $r(t)$ as defined in \Fig{front_def}.
%Thin line: Power-law fit to the front position, $r(t) \propto t^a$.
}
\end{figure*}

In \Fig{Jz0_Fig} and \ref{Jz_Fig}, we visualize the time evolution of
the local magnetization in density plots,
with site index $n$ on the $x$-axis, time $t$ on the $y$-axis.
Here, the absolute value of the magnetization is shown as a
grayscale and in lines of constant magnetization at $|\aver{S_z}| =$ 0.2,
0.4.
In \Fig{Jz0_Fig}, the relation between the density plots and the actual
magnetization profile for the XX model is shown at two times,
$t=0$ and $t=40$.
The exact solution is perfectly 
reproduced, including the detailed substructure of the magnetization
profile.

In \Fig{Jz_Fig}, density plots for various values of $J_z$ between $0$
and $1.1$ are shown. 
For small $J_z$ ($J_z<1$), we observe ballistic transport of the
magnetization. This regime is characterized by a constant transport
velocity of the magnetization, hence the lines of constant
magnetization shown in \Fig{Jz_Fig} are approximately straight
for $J_z<1$.
The magnetization front propagation slows down as 
$J_z$ increases, and almost comes to a halt when $J_z > 1$.
Although the sharpness of this crossover at $J_z=1$ is surprising,
its general nature can be understood from the limits
$J_z \to 0$ and $|J_z| \to \infty$:
For small $J_z \to 0$ the
$S^xS^x$- and $S^yS^y$-interactions dominate. Being spin flip terms,
they smear out the initially hard step profile in the $z$
magnetization. 
For large $J_z$, on the other hand, the $S^zS^z$-interaction dominates.
This term does not delocalize the step profile, and in the limit
$\abs{J_z} \to \infty$, the initial state is even a stationary
eigenstate of the Hamiltonian.

Besides the structure of the overall front, we also observe for $J_z
\neq 0$ remnants of the steplike substructure from the XX model, individual
pockets of transported magnetization at velocity 1, which we
call ``carriers''. 
As $J_z$ is increased, these carriers keep the velocity $v \approx 1$, 
but are increasingly damped and thus less and less effective in
transporting magnetization.

\begin{figure}
\epsfig{file=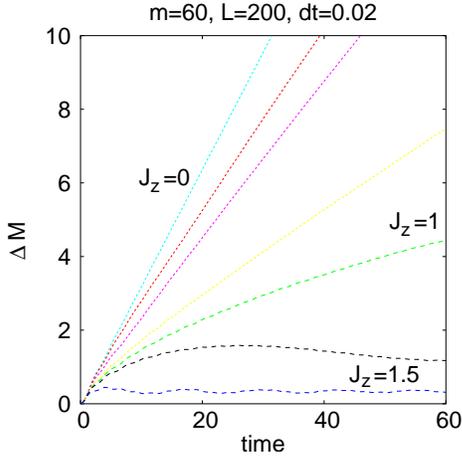, width=.7\linewidth}
\caption[The change in the magnetization $\Delta M (t)$ for
$J_z=0;\;0.3;\;0.6;\; 0.9;\; 1.0;\; 1.1;\; 1.5$]{ 
  \label{totalmagnJz}
  The change in the magnetization 
  $\Delta M (t)$ is shown. The curves are plotted in the order
  $J_z=0;\;0.3;\;0.6;\; 0.9;\; 1.0;\; 1.1;\; 1.5$, where $J_z=0$ is
  the steepest. The curves $J_z=0;\;0.3;\;0.6;\; 0.9$ show the same
  linear behaviour for the observed times, i.e.\ up to $t=60$.
}
\end{figure}

In order to put the above observations on a more quantitative footing,
we plot in \Fig{totalmagnJz} the integrated flow of magnetization
through the center, 
\be
\Delta M(t) 
= \int_0^t \aver{j_{L/2}(t')} \textrm{d}t'
= \sum_{n>L/2}^L (\aver{S^z_n(t)}+1/2).
\ee 
%and in \Fig{totalcurrentJz} the total current $\aver{J(t)}$. 
This quantity has the advantage that unlike the lines of constant
 magnetization in \Fig{Jz0_Fig} and \ref{Jz_Fig},  it shows the
overall spin transport without being too much biased by single
``carriers''.

We observe in \Fig{totalmagnJz} roughly linear behaviour of $\Delta M(t)$ for
$\abs{J_z}<1$, which suggests ballistic magnetization  
transport at least on the time scales accessible to us. As $J_z$ increases, 
magnetization transport slows down
until around $J_z=1$ the behaviour changes drastically:
For $J_z>1$, $\Delta M(t)$ seems to saturate at a finite value, around
which it oscillates.
On the
time scales accessible to us, we thus find a sharp crossover at  
$J_z=1$ from ballistic transport to an almost constant magnetization.
%, with superdiffusive magnetization transport at $J_z=1$. 

\begin{figure}
\epsfig{file=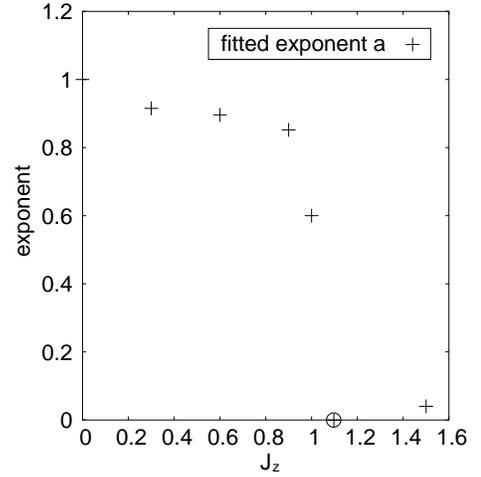, width=.7\linewidth}
\caption[Best fit for the exponent $a$ in $\Delta M(t) \propto t^a$.]{ \label{exponentJz}
Best fit for the exponent $a$ in $\Delta M(t) \propto t^a$, for the
data shown in \Fig{totalmagnJz} and for times between $t=20$ and
$t=60$.
We estimate the uncertainty in $a$ to be of the order of 0.1 due to
the limited time available (cf.\ \Fig{Jz1}).
It was not possible to fit the slow oscillations for $J_z=1.1$.
To the eye, however, the curve in \Fig{totalmagnJz} suggests slow
oscillations around a constant value, hence we included in the data
point $a=0$ for $J_z=1.1$ by hand (encircled).
}
\end{figure}

This crossover is even more clearly visible in \Fig{exponentJz},
where we plot the exponent $a$ of the magnetization, $\Delta M(t)
\propto t^a$, for values $J_z$ between 0 and 1.5. 
Here, the exponent $a$ is close to 1 for $J_z<1$, confirming the
roughly linear transport, and quickly drops to zero in the regime of
constant magnetization for $J_z>1$. 

\begin{figure}
\epsfig{file=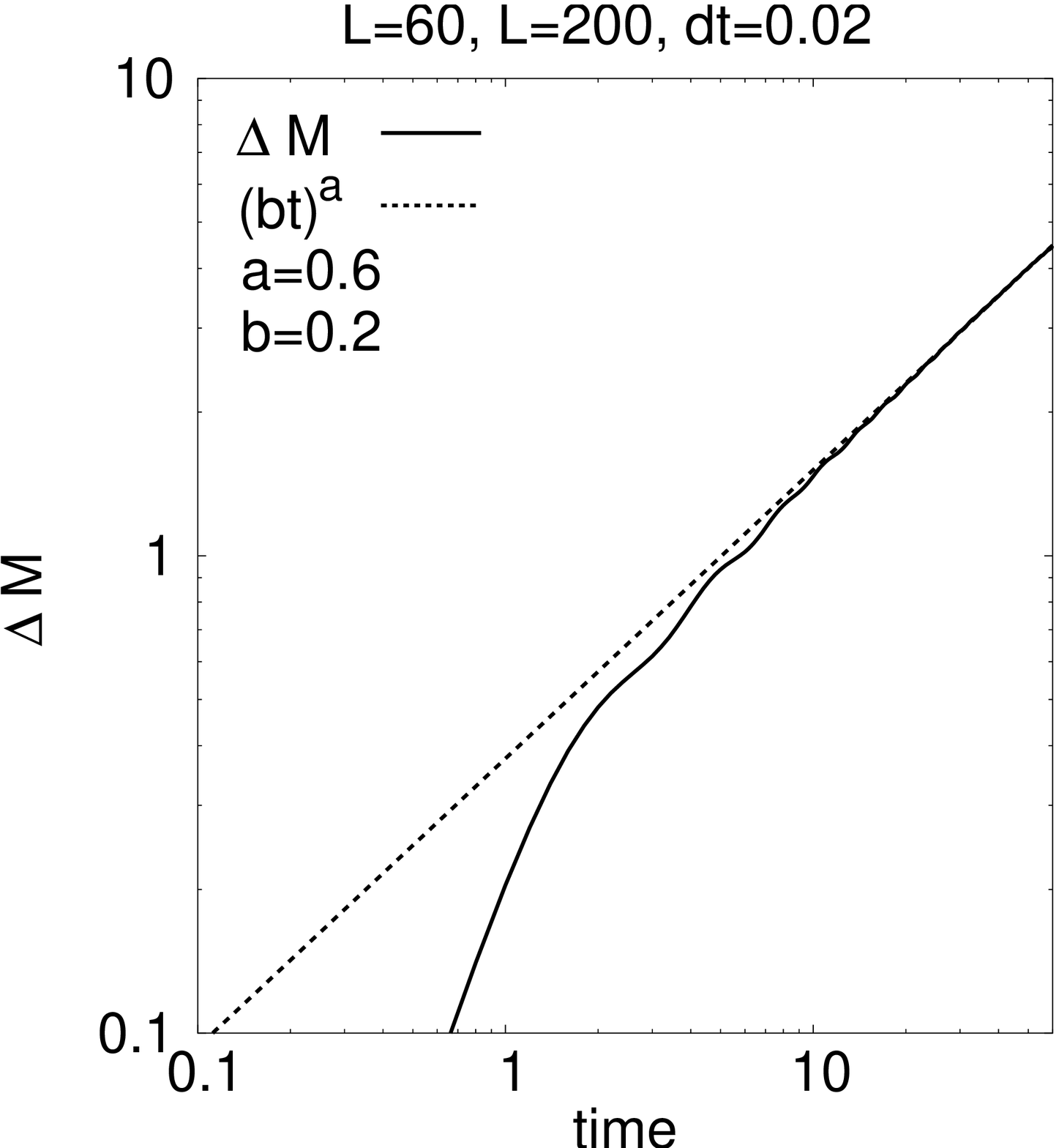, width=.7\linewidth}
\caption{ \label{Jz1} $J_z=1$: 
The change of the magnetization in a double logarithmic plot with
 an algebraic fit.
}
\end{figure}

\begin{figure}
\epsfig{file=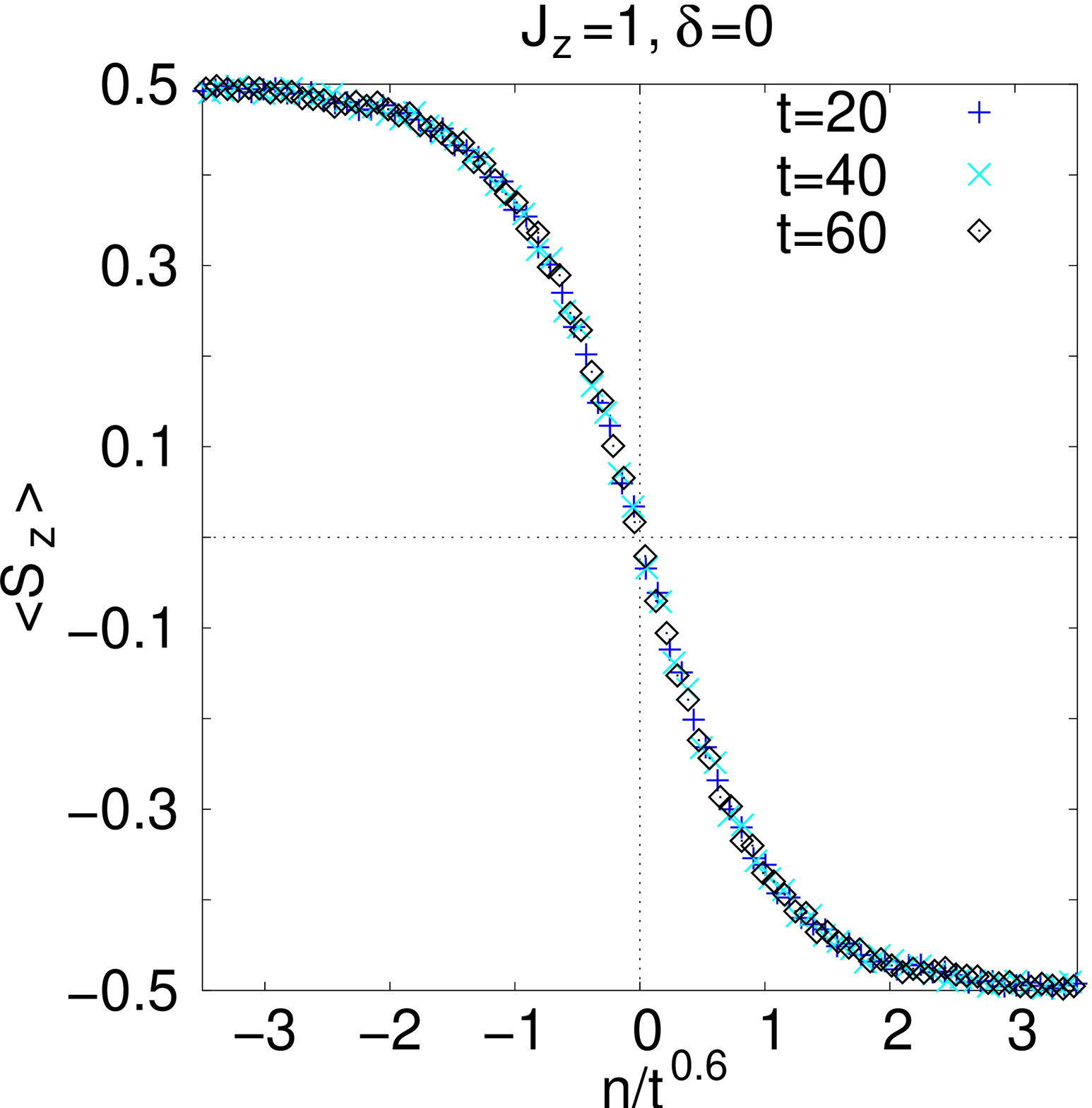, width=.6\linewidth}
\caption{ \label{Jz1scal} $J_z=1$: Collapse of magnetization for a 
superdiffusive scaling form $(x/t^{0.6})$.
}
\end{figure}

\Fig{Jz1} illustrates how the exponent $a$ was obtained, for the
special case $J_z=1$.
Here the exponent $a = 0.6 \pm 0.1$ indicates that
the magnetization transport is clearly not
ballistic anymore. In fact, 
we find from a scaling plot \Fig{Jz1scal} that for long times the
magnetization collapses best for a scaling function of the form 
$S_z(n,t)\sim\phi(n/t^{0.6})$ with an uncertainty in the exponent of approximately $0.1$, indicating superdiffusive or diffusive transport in the time range under consideration.

The proposed crossover from ballistic to almost no transport is
also visible in the expectation value of the current $j_n = J_n \Im
(\aver{ S_n^+ S_{n+1}^-0})$. 
For $J_z=\delta=0$, it is known \cite{AntalSchutz99} that the
current at the middle of the chain approaches a finite value as $t
\rightarrow \infty$.
This is only possible for ballistic transport. 
In the case of (sub- / super-) diffusive transport or constant / oscillatory
magnetization, on the other hand, the central 
current must fall off to zero as the magnetization gradient flattens
or must even become negative to allow for the oscillations.

\begin{figure}
\epsfig{file=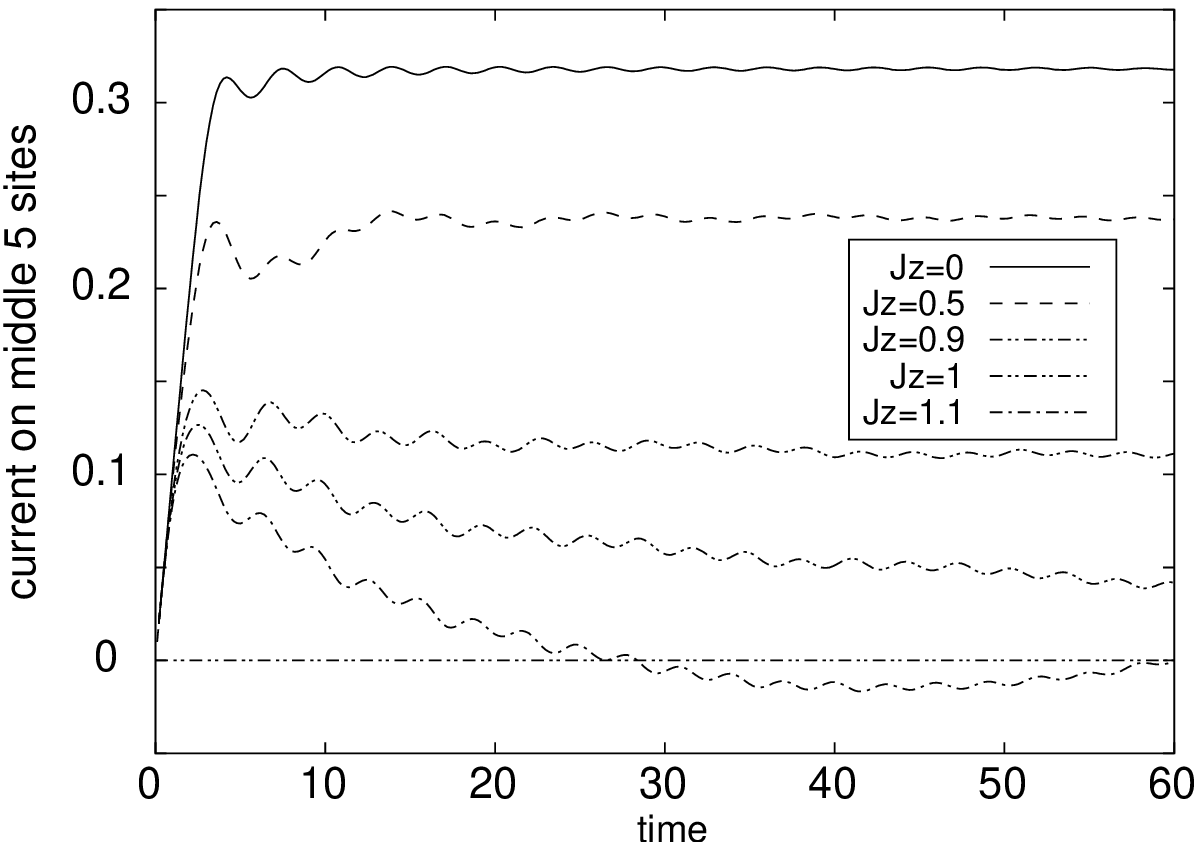, width = 1\linewidth}
\caption{ \label{curr_scal_Jz}
Current, averaged over the 5 middle sites, for various values of $J_z$
between $0$ and $1.1$. 
}
\end{figure}
This expected behaviour is seen in \Fig{curr_scal_Jz}, 
where we plot the current at the center of the chain
as a function of time for various values of $J_z$ between $0$ and $1.1$.
We averaged the current over the 5 middle sites in order to filter out
local current oscillations. 
%For the given parameters, the runaway time
%was determined by a convergence analysis in $m$ to be at $t_R = ...$.
We observe that for relatively long times, the current approaches
a constant value for $|J_z|<1$, whereas the current falls off rapidly and 
then seems to exhibit damped oscillations around zero for $|J_z|>1$.
This strengthens our previous conclusion of a crossover from ballistic
transport to a more or less constant magnetization at $|J_z| = 1$.

Remarkably, this crossover for the behaviour of a high-energy quantum
state $\ketini$
is found at the location $J_z=1$ of the quantum phase transition from the critical 
phase to the N\'eel antiferromagnetic state (see \Fig{phasediagram}),
a priori a low-energy event. To understand the subtle connection
between the time evolution of $\ketini$ and the phase transition,
we exploit that the 
time-evolution does not depend on the sign of $J_z$, as discussed in
Sec.\ \ref{time_model}.
Therefore the time evolution of the high-energy state $\ketini$ for
$J_z >1$ is identical to that for $J_z' = -J_z <-1$, where $\ketini$ is a
low-energy state.
At the quantum phase transition from the ferromagnetic
state to the critical phase at $J_z'=-1$ the ground state, 
a kink state for $J_z'<-1$ (if we impose the boundary condition spin
up on the left boundary and spin down on the right boundary) \cite{Mikeska91},
changes drastically to a state with no kink and power-law correlations
for $J_z'>-1$.
Therefore, our initial state is very close to an eigenstate -- the ground
state -- for $J_z'<-1$, but not for $J_z' > -1$.
Thus,  the harsh change in the time-evolution of the high-energy
state $\ket{\init}$ at $J_z=1$ can be explained by the severe change
in the ground state properties at $J_z'=-1$, and
the crossover is linked to a quantum phase transition at a
different location in the phase diagram.  

We now study the influence of a nonzero dimerization $\delta$ in
\Eq{Hamilton}. We restrict our analysis to the case $J_z=0$.
The dimerized models can still be described in terms of the
free-fermion picture and are exactly solvable (for static properties
see \olcite{Schuetz94}).
The current, however, is not conserved for nonzero dimerization.
This example will shed light on the question whether the long-time limit
depends on current conservation or on the free-fermion property, or
yet on other  special properties of the system.
We expect two obvious effects of nonzero dimerization: 
Firstly, the overall front velocity should slow down, because
the magnetization now propagates faster  on half of the links, but slower 
on the other half, the net effect being a reduction of the total
velocity.
Secondly, we expect oscillations with a period of two lattice sites.
This is obvious in the limit $\delta \rightarrow 1$, where each
strongly coupled pair of sites can be viewed as an almost isolated
subsystem, in which the magnetization oscillates back and forth.
We expect remnants of this behaviour also at dimerizations
$|\delta| < 1$.
%For $\delta$ close to 1, we expect a (boring) crossover to weakly
%coupled two-level systems.

\begin{figure*}
\epsfig{file=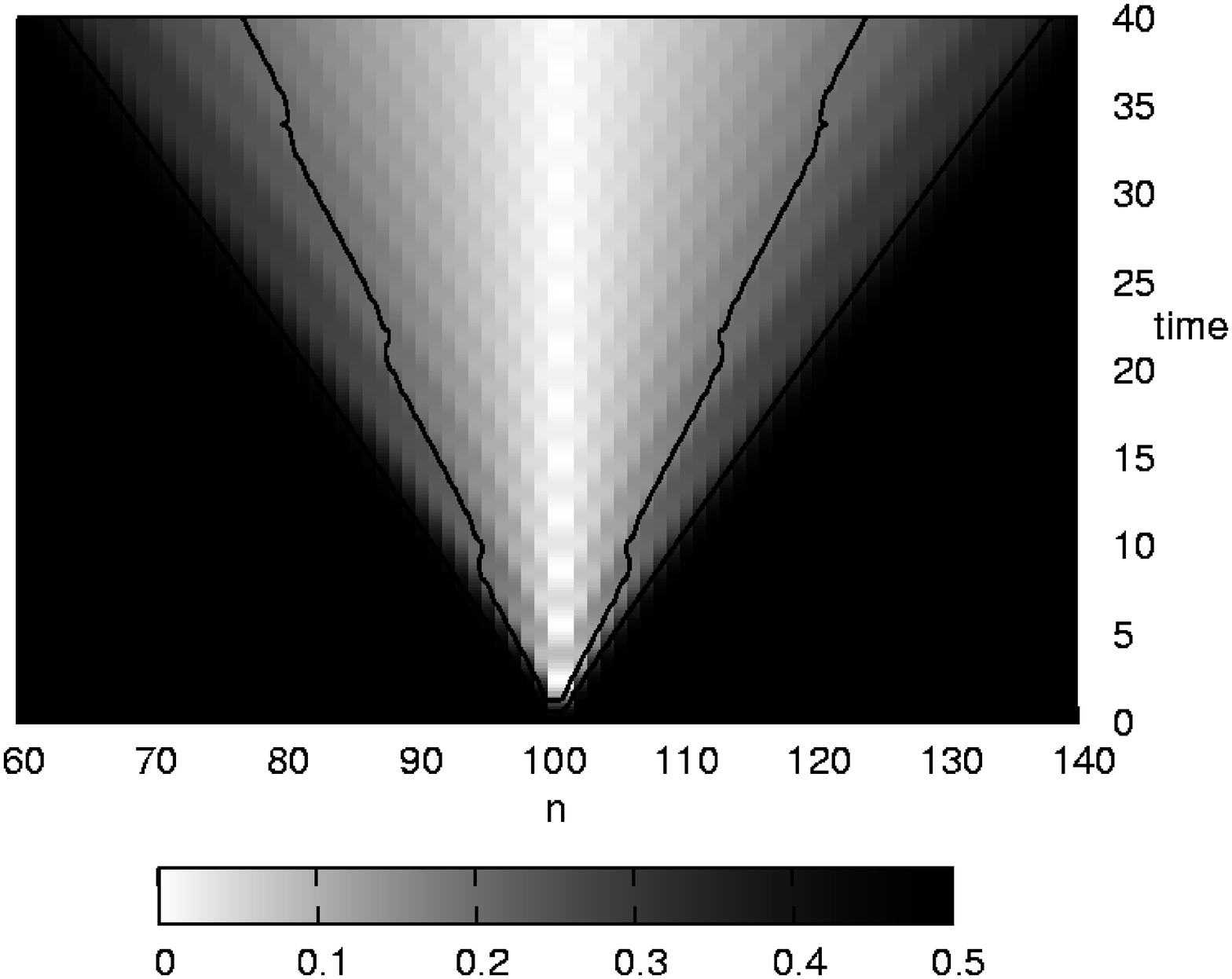,  width=.32\linewidth}
\epsfig{file=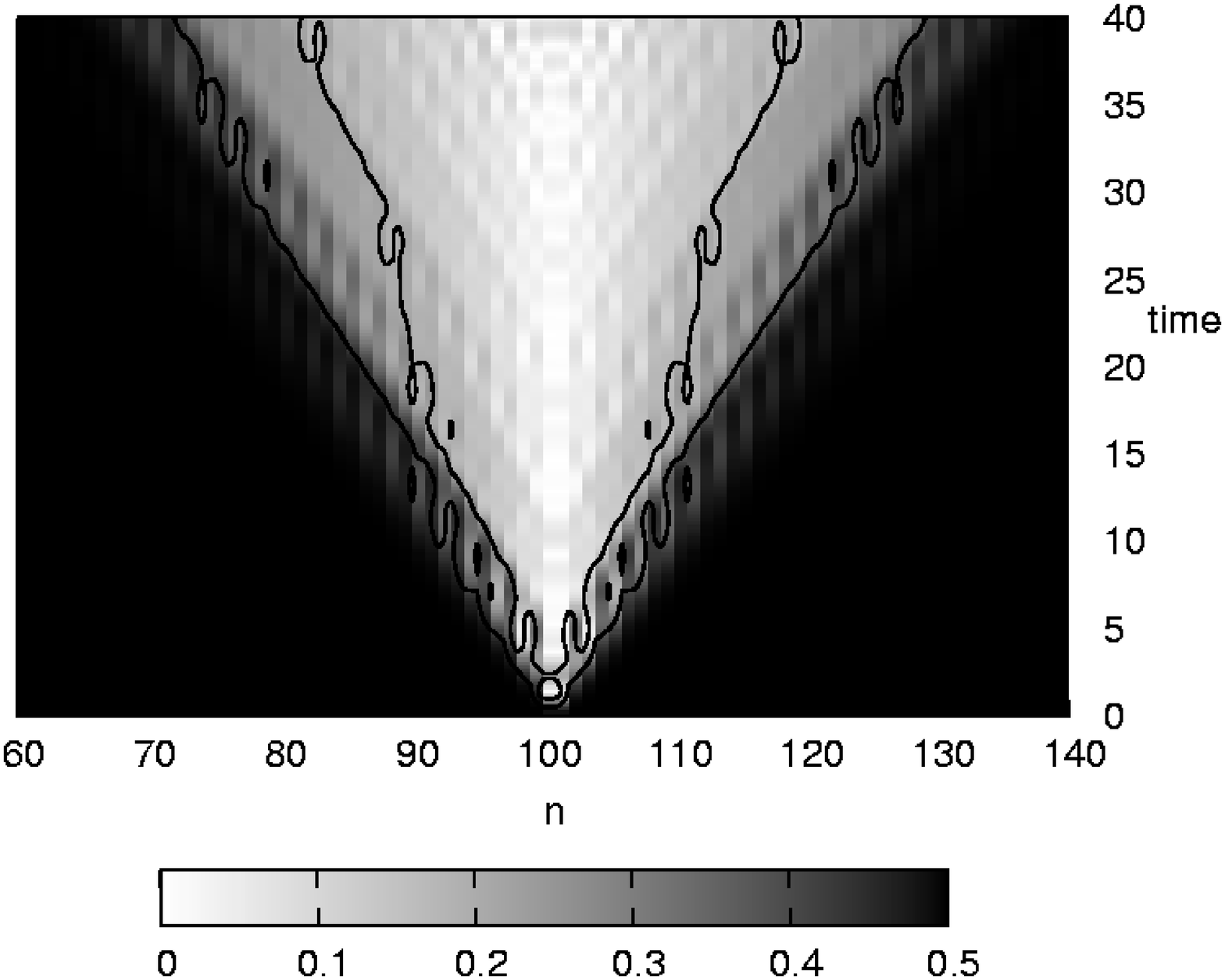,  width=.32\linewidth}
\epsfig{file=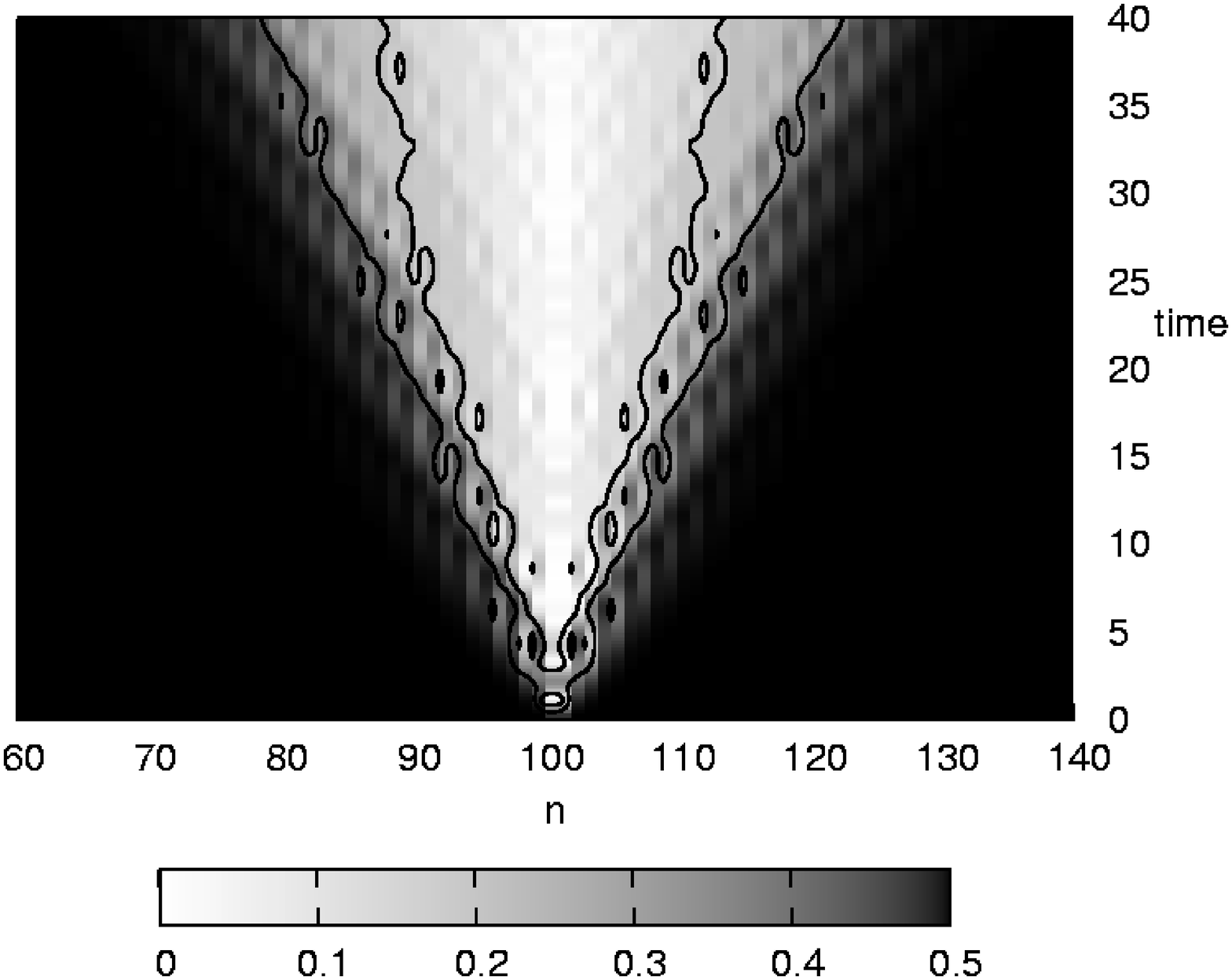,  width=.32\linewidth}
\epsfig{file=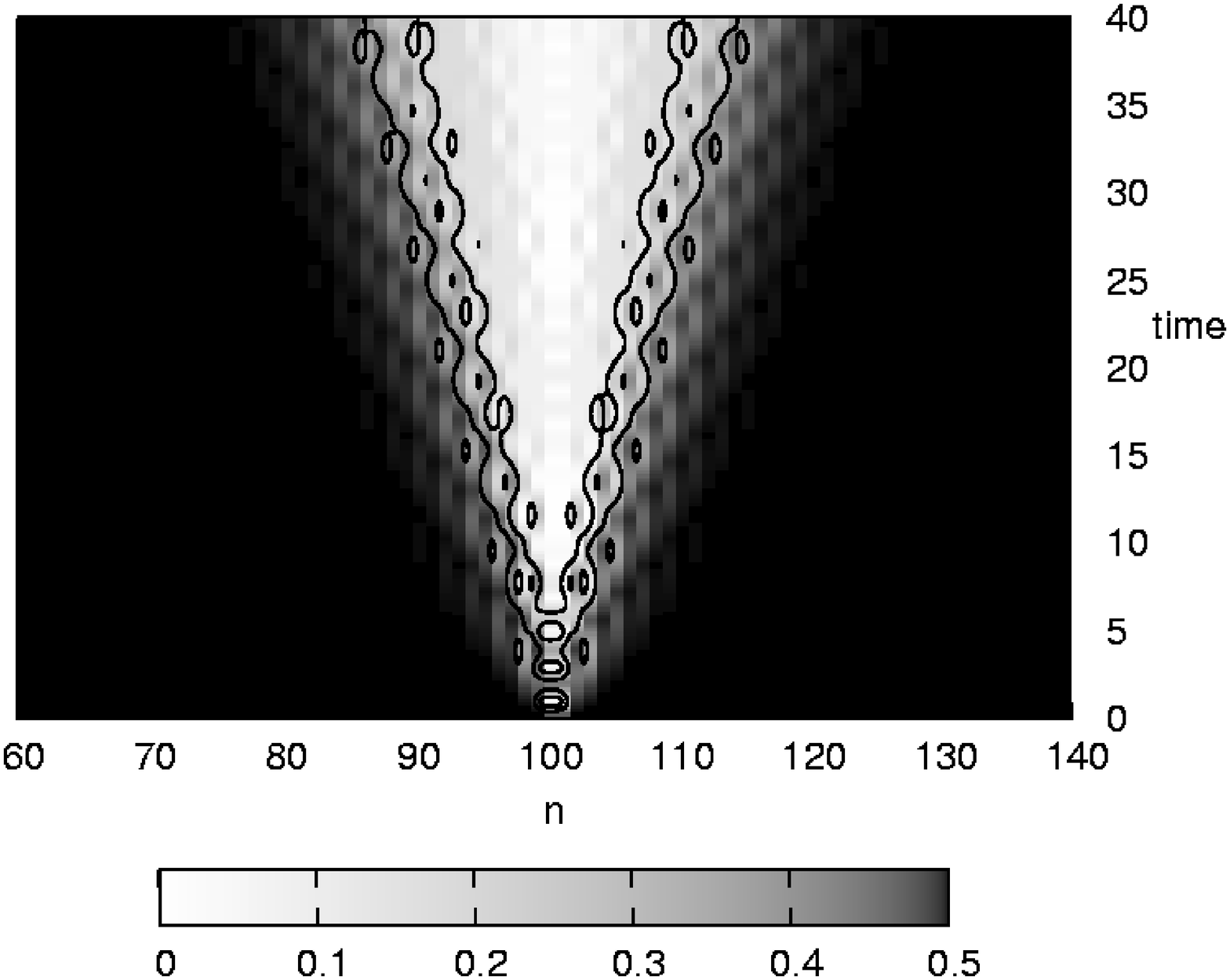,  width=.32\linewidth}
\epsfig{file=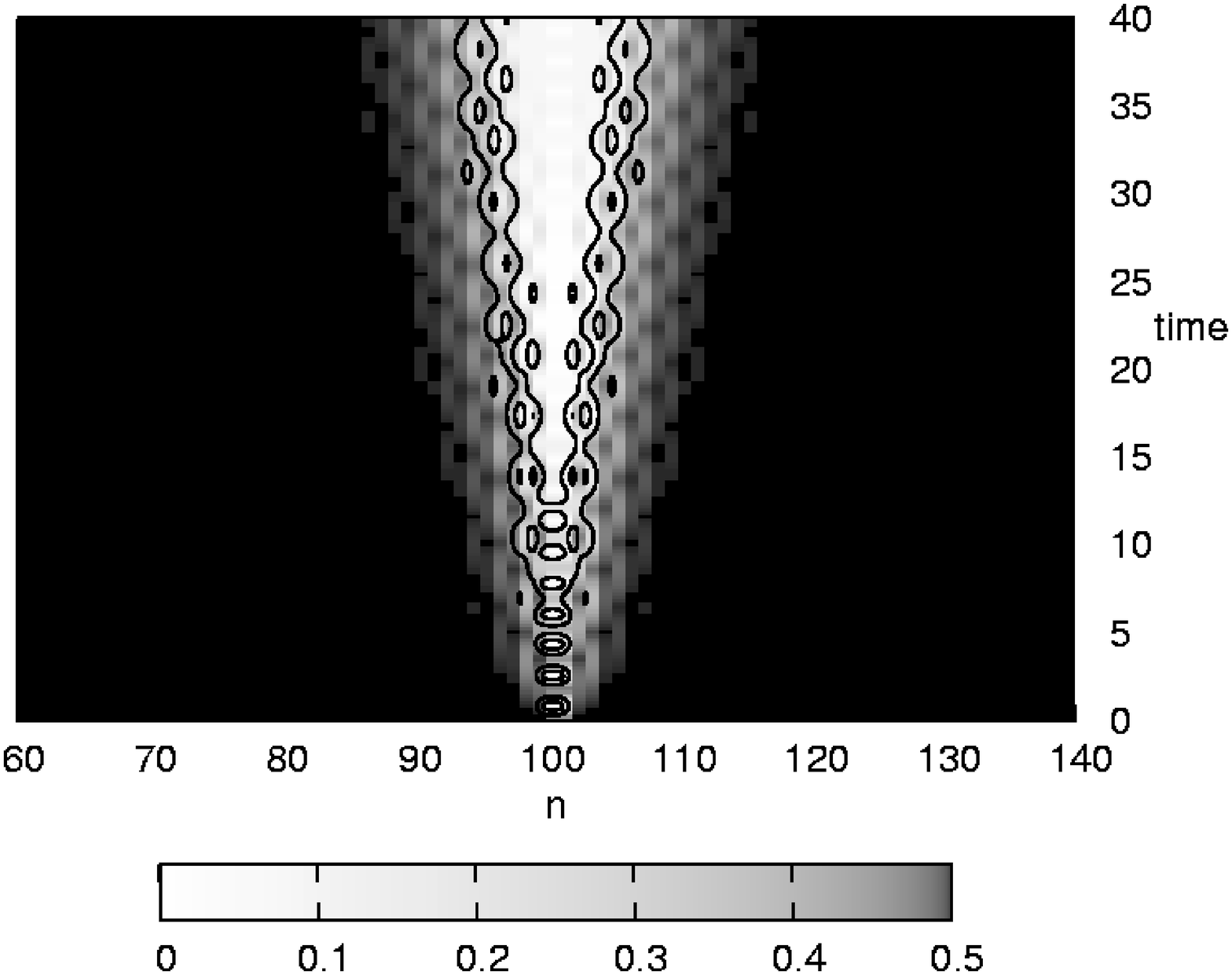,  width=.32\linewidth}
\epsfig{file=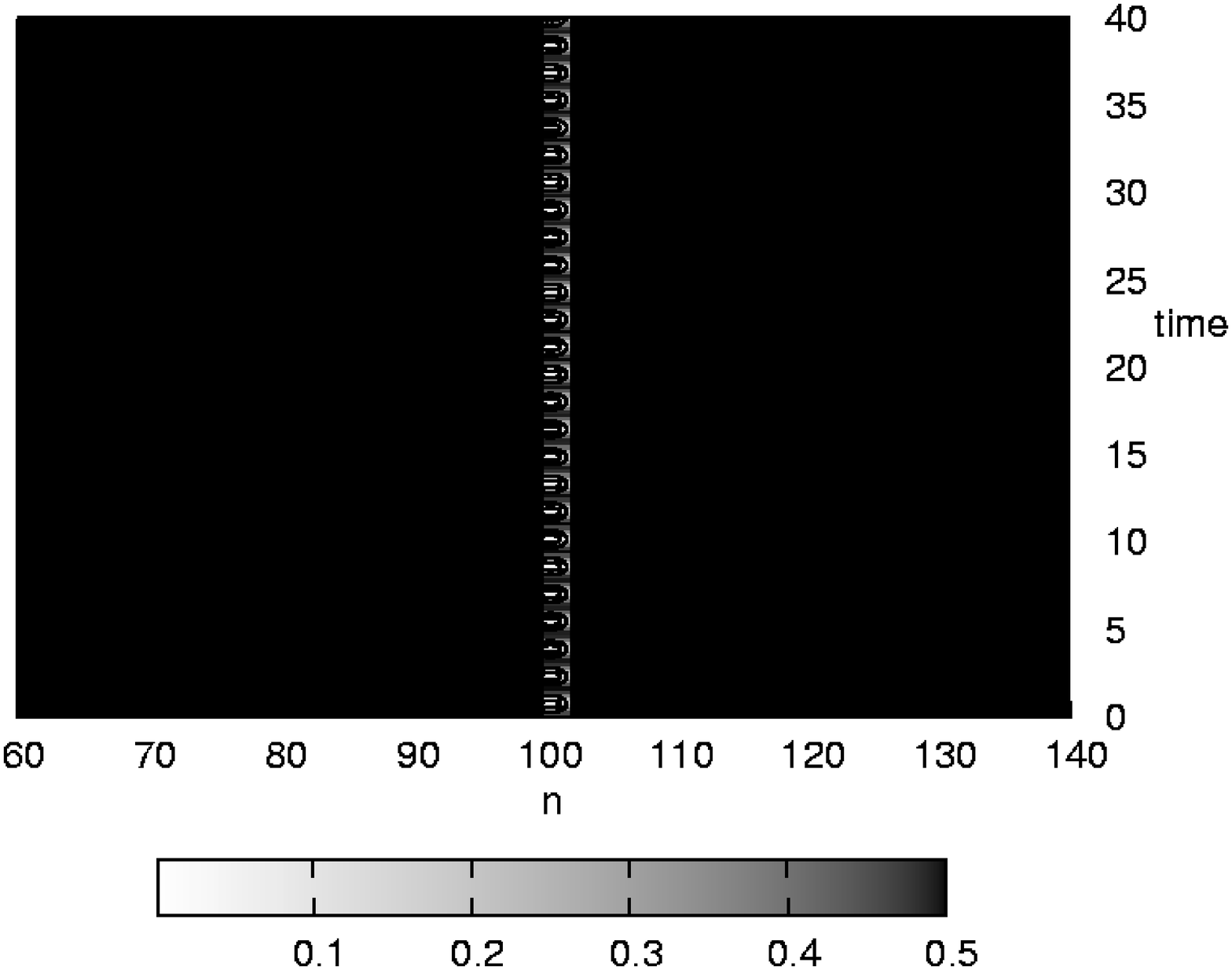,  width=.32\linewidth}
\caption[Density plots of the magnetization $\langle S^z_n(t) \rangle$
for dimerization $\delta=0;\;0.2;\; 0.4;\; 0.6;\; 0.8;\; 1.0$, and
$J_z = 0$.]{ 
\label{dimer_L200} 
Density plots of the magnetization $\langle S^z_n(t) \rangle$ as in \Fig{Jz_Fig},
for dimerization (from left to right, top to bottom) $\delta=0; 0.2; 0.4;
0.6; 0.8; 1.0$, and $J_z = 0$. 
The grayscale mapping is different in each plot as indicated by the legends.
Solid lines: lines of constant magnetization $\langle S^z_n \rangle =
\pm 0.2, \pm 0.4$.
}
\end{figure*}

The data shown in \Fig{dimer_L200} confirms this
expectation qualitatively, but does not reveal any other qualitative
change of the long-time limit for nonzero dimerization.
\begin{figure}
\epsfig{file=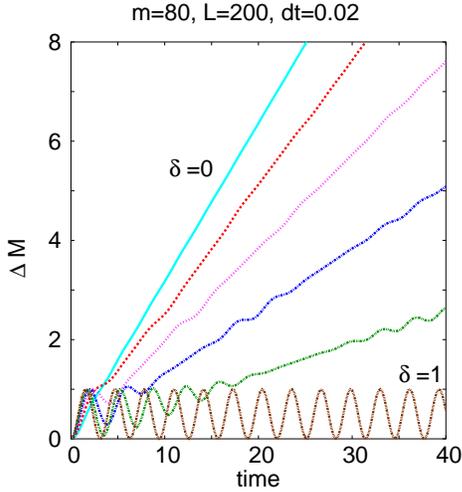, width=.7\linewidth}
\caption{ \label{totalmagnd}
Change in magnetization $\Delta M(t)$ for different dimerizations,
from top to bottom: $\delta =$ 0, 0.2, 0.4, 0.6, 0.8, 1.0.
}
\end{figure}
\begin{figure}
\epsfig{file=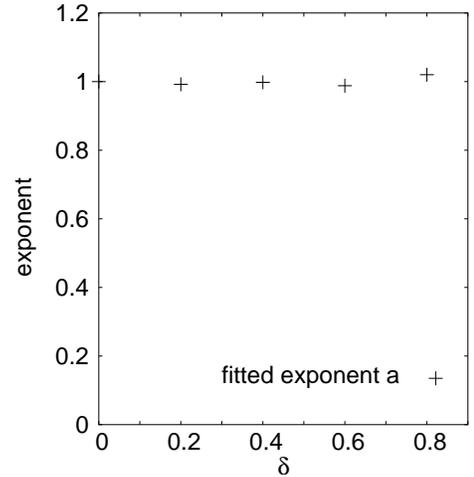, width=.7\linewidth}
\caption[Best fit for the exponent $a$ in $\Delta M(t) \propto t^a$.]{ \label{exponentd}
Best fit for the exponent $a$ in $\Delta M(t) \propto t^a$, for the
data shown in \Fig{totalmagnJz} and for times between $t=20$ and
$t=40$.
}
\end{figure}
For $\delta = 1$, the system is trivially given by isolated pairs of
neighbouring sites, therefore the propagation velocity drops to zero.

\Fig{totalmagnd} and \Fig{exponentd} reveals explicitly that no
qualitative change occurs as the dimerization is switched on:
the change in magnetization $\Delta M(t)$ still shows the linear
behaviour typical of ballistic transport.
For increasing $\delta\to 1$ oscillations on top of this
linear behaviour arise. 
We find that switching on finite 
dimerization does not change the long-time behaviour of the time
evolution also for nonzero $J_z$ (not shown).
In particular, the time evolution here is drastically influenced by the
transition at $J_z = 1$ as in the case $\delta = 0$ discussed above.

To summarize, we find the same long-time behaviour of the initial
state $\ini$ in the dimerized system --- a system with gapped excitation
spectrum and which is
exactly solvable--- as in the system with small $S^zS^z$-interaction,
$\abs{J_z}<1$ --- a system which is critical --- whereas the behaviour
changes drastically for larger $S^zS^z$-interaction,
$\abs{J_z}>1$. 
Hence we cannot attribute the ballistic transport of the
magnetization to the specific properties of the XX model; 
neither to be exactly solvable, nor
to the continuous spectrum nor to the conserved current in the XX-model.
The drastic change at $\abs{J_z}=1$ stems from the special property of the
initial state to resemble the ground state in the ferromagnetic phase
and the highest energy state in the antiferromagnetic phase.

%\begin{figure}
%\epsfig{file=../curr_scal/current_average_Jz0_d0.eps, width=.45\linewidth}
%\epsfig{file=../curr_scal/current_average_Jz0_d0.2.eps, width=.45\linewidth}
%\epsfig{file=../curr_scal/current_average_Jz0_d0.4.eps, width=.45\linewidth}
%\epsfig{file=../curr_scal/current_average_Jz0_d0.6.eps, width=.45\linewidth}
%\epsfig{file=../curr_scal/current_average_Jz0_d0.8.eps, width=.45\linewidth}
%\epsfig{file=../curr_scal/current_average_Jz0_d1.eps, width=.45\linewidth}
%\caption{ \label{dimercurrent}
%Current, averaged over the 5 middle sites, for $J_z=0$ and various values of
%dimerization $\delta$ between $0$ and $1$.  
%TODO: How does it look for a single site? 
%For $\delta=1$, only the spin at the two central sites
%oscillates back and forth. It is uncoupled from the remainder of the
%chain, in which the aligned spins of the initial state stay constant.
%}
%\end{figure}

Finally, let us include a note on the errors in the present analysis.
A convergence analysis in $m$ as in section \ref{time_error}
shows that the errors and the runaway time are roughly the same as
for the XX model. The plot in \Fig{Jz_Fig} goes up to time $t=95$, 
whereas the runaway
time $t_R$ is somewhat earlier, $t_R \approx 60-80$, depending on the
precise value of $J_z$.  
Indeed, a convergence analysis in $m$ reveals that the accuracy
in the central region decreases for $t>t_R$.
%This central region is, however, small enough for the scaling analysis
%not to be seriously affected. 
For dimerized models the runaway time 
$t_R$ is somewhat shorter (between $t_R=40$ and $t_R=80$ for $m=50$, 
depending on the
dimerization). This fact reflects the reduced accuracy 
of the DMRG algorithm when  
dealing with inhomogenous systems. 
As always, it is possible to increase $t_R$ by increasing $m$. 

\section{Conclusions}
We have investigated the evolution of the initial state $\ini$ under the 
effect of nearest-neighbour interactions with the adaptive time-dependent DMRG. 

For weak $S^zS^z$-interaction,
i.e.\ $\abs{J_z}<1$  in Eq. (\ref{Hamilton}), and arbitrary
dimerization, $0\le \delta<1$, we find that for long times the
transport of the magnetization is ballistic as it was found for the
$XX$-model\cite{AntalSchutz99}. The magnetization profile
shows the same scaling form for long times,
i.e.\ $S^z(n,t)=\varphi((n-n_c)/t)$, where $n_c$ is the position of
the chain center, but with different
scaling functions $\varphi$. For stronger
$S^zS^z$-interaction, i.e. $\abs{J_z}>1$, even in a homogeneous
system, $\delta=0$, a drastic change in the long-time evolution is
seen. The magnetization transport is no longer ballistic, but shows
oscillatory behaviour around a constant value. Hence our results suggest
that the specific properties of the XX model are not responsible for ballistic
transport at long times. The drastic change in the long time behaviour
at the phase transition $J_z=1$ can be attributed to the close resemblance 
of the initial state to the ground state for $J_z<-1$.

Our error analysis for the adaptive time-dependent DMRG shows 
that for small times the error is
dominated by the Trotter error whereas for long times the truncation
error becomes the most important. 
This finding should be general and hold for non-exactly solvable
models as well, and should therefore allow to control
the accuracy of the results of adaptive time-dependent DMRG
in general models. Overall, we find this method to be very precise at
relatively long times.

{\em Acknowledgments.} US wishes to thank the Aspen Center for Physics,
where parts of this work were completed, for its hospitality. The authors are 
grateful for discussions with Joel Lebowitz, Herbert Spohn, Hans-J\"{u}rgen
Mikeska, Attila Rakos, Ian McCulloch, Zoltan R\'acz and 
Vladislav Popkov. CK and US acknowledge support by the 
Studienstiftung des deutschen Volkes and the Young Academy, Berlin, 
respectively.

\end{document}